\pdfoutput=1
\documentclass[conference]{IEEEtran}
\usepackage{hyperref}
\newif\ifanon

\ifanon\else%
  \IEEEoverridecommandlockouts%
\fi

\usepackage{ebproof}
\usepackage{booktabs}
\usepackage{jms-sections}
\usepackage{stmaryrd}
\usepackage{local-jmsdelim}
\usepackage{local-abbrv}
\usepackage{cite}
\usepackage{microtype}
\usepackage{amsmath,amssymb,amsfonts}
\usepackage{algorithmic}
\usepackage{graphicx}
\usepackage{textcomp}
\usepackage{xcolor}
\usepackage{url}
\usepackage{tikz-cd}
\usepackage{mathtools}

\DeclareFontFamily{OMX}{MnSymbolE}{}
\DeclareSymbolFont{mnomx}{OMX}{MnSymbolE}{m}{n}
\SetSymbolFont{mnomx}{bold}{OMX}{MnSymbolE}{b}{n}
\DeclareFontShape{OMX}{MnSymbolE}{m}{n}{
    <-6>  MnSymbolE5
   <6-7>  MnSymbolE6
   <7-8>  MnSymbolE7
   <8-9>  MnSymbolE8
   <9-10> MnSymbolE9
  <10-12> MnSymbolE10
  <12->   MnSymbolE12}{}

\makeatletter
\def\Decl@Mn@Delim#1#2#3#4{%
  \if\relax\noexpand#1%
    \let#1\undefined
  \fi
  \DeclareMathDelimiter{#1}{#2}{#3}{#4}{#3}{#4}}
\def\Decl@Mn@Open#1#2#3{\Decl@Mn@Delim{#1}{\mathopen}{#2}{#3}}
\def\Decl@Mn@Close#1#2#3{\Decl@Mn@Delim{#1}{\mathclose}{#2}{#3}}
\Decl@Mn@Open{\llangle}{mnomx}{'164}
\Decl@Mn@Close{\rrangle}{mnomx}{'171}
\Decl@Mn@Open{\lmoustache}{mnomx}{'245}
\Decl@Mn@Close{\rmoustache}{mnomx}{'244}
\makeatother

\usetikzlibrary{graphs,decorations.pathmorphing,decorations.markings}

\def\BibTeX{{\rm B\kern-.05em{\sc i\kern-.025em b}\kern-.08em
    T\kern-.1667em\lower.7ex\hbox{E}\kern-.125emX}}
\begin{document}

\allowdisplaybreaks

\newtheorem{theorem}{Theorem}[section]
\newtheorem{notation}[theorem]{Notation}
\newtheorem{observation}[theorem]{Observation}
\newtheorem{definition}[theorem]{Definition}
\newtheorem{remark}[theorem]{Remark}
\newtheorem{lemma}[theorem]{Lemma}
\newtheorem{corollary}[theorem]{Corollary}
\newtheorem{convention}[theorem]{Convention}

\NewDocumentCommand\WTp{m}{\mathbf{W}\Sub{#1}}
\NewDocumentCommand\MTp{m}{\mathbf{M}\Sub{#1}}
\NewDocumentCommand\LSegal{}{\mathcal{L}_{\textit{horn}}}
\NewDocumentCommand\LRezk{}{\mathcal{L}_{\textit{biinv}}}
\NewDocumentCommand\LComplete{}{\mathcal{L}_{\textit{cmpl}}}
\NewDocumentCommand\LScn{}{\mathcal{L}_{\textit{scn}}}

\NewDocumentCommand\Type{}{\mathsf{Type}}
\NewDocumentCommand\Poly{m}{\mathbf{P}\Sub{#1}}
\NewDocumentCommand\II{}{\mathbb{I}}
\NewDocumentCommand\ICh{m}{\Delta^{#1}}
\NewDocumentCommand\Empty{}{\mathbf{0}}
\NewDocumentCommand\One{}{\mathbf{1}}
\NewDocumentCommand\Two{}{\mathbf{2}}
\NewDocumentCommand\Three{}{\mathbf{3}}
\NewDocumentCommand\WalkingBiinv{}{\mathbb{E}}
\NewDocumentCommand\Horn{}{\Lambda^{\Two}_{1}}
\NewDocumentCommand\Prop{}{\Omega}
\NewDocumentCommand\Lam{m}{\lambda{#1}\mathpunct{.}}
\NewDocumentCommand\IsT{m}{\bbrk{#1}}
\NewDocumentCommand\IsF{m}{\ggl{#1}}

\NewDocumentCommand\OpenEmbeddings{}{\mathcal{O}}
\NewDocumentCommand\OpenProp{}{\Prop\Sub{\OpenEmbeddings}}

\title{When is the partial map classifier a Sierpi\'nski cone?\ifanon\else
\thanks{This work was funded by the United States Air Force Office of Scientific Research under grant FA9550-23-1-0728 (\emph{New Spaces for Denotational Semantics}; Dr Tristan Nguyen, Program Manager). Views and opinions expressed are however those of the authors only and do not necessarily reflect those of AFOSR.}\fi
}

\author{
  \ifanon
  \IEEEauthorblockN{Anonymous Authors}
  \else
  \IEEEauthorblockN{Leoni Pugh}
  \IEEEauthorblockA{%
    \textit{Computer Laboratory} \\
    \textit{University of Cambridge}\\
    Cambridge, UK\\%
    \url{lxp20@cantab.ac.uk}
  }
  \and
  \IEEEauthorblockN{Jonathan Sterling}
  \IEEEauthorblockA{%
    \textit{Computer Laboratory} \\
    \textit{University of Cambridge}\\
    Cambridge, UK\\%
    \url{js2878@cl.cam.ac.uk}
  }
  \fi
}

\maketitle
\thispagestyle{plain}
\pagestyle{plain}

\begin{abstract}

  We study the relationship between partial map classifiers, Sierpi\'nski cones, and axioms for synthetic higher categories and domains within univalent foundations. In particular, we show that synthetic $\infty$-categories are closed under partial map classifiers assuming Phoa's principle, and we isolate a new reflective subuniverse of types within which the Sierpi\'nski cone (a lax colimit) can be computed as a partial map classifier by strengthening the Segal condition.
\end{abstract}


\begin{xsect}{Introduction}
  \begin{xsect}{Axiomatic and synthetic approaches to mathematics}
    It usually happens that a given mathematical concept can be fruitfully studied in terms of the (possibly higher) \emph{category} of its instances and their structure-preserving homomorphisms. This is true of all algebraic structures like groups and rings and categories themselves, as well as any kind of space—including \emph{domains}, the elusive spaces that computer scientists use to develop the denotational semantics of programming languages.

    Practitioners of the \emph{axiomatic method} tend rather quickly to replace this specific category by an arbitrary category satisfying some axioms that reflect the decisive properties of the original one. For example, axiomatic category theory takes place in an arbitrary 2-category with enough structure, such as a cosmos or some kind of (2,2)-topos; likewise, axiomatic domain theory~\cite{fiore:1994} takes place in a cartesian closed category equipped with a monad satisfying certain axioms.

    A refinement of the axiomatic method is the \emph{synthetic method} which today focusses specifically on those axiomatic models that can be obtained as \emph{full subcategories} of an ($n$,1)-topos, or to be more precise, full subfibrations of an ($n$,1)-topos. The case of category theory itself is illustrative: axiomatic $n$-category theory could take place in an ($n+1$,$n+1$)-topos, whereas synthetic $n$-category theory would take place in a full subfibration of an ($n+1$,1)-topos.

    The reason to develop synthetic models of mathematical concepts is to make continuity conditions disappear: for example, a synthetic category or a synthetic domain is nothing more than a \emph{type} satisfying some simple laws with no additional structure, and any function whatsoever satisfies the necessary functoriality or continuity law automatically. In return, we must only use reasoning that is valid in the topos---so we must usually give up the full axiom of choice, \etc.
    In the current era in which proof assistants based on Martin-L\"of type theory flourish~\cite{norell:2009,coq:reference-manual,lean:2015}, the synthetic approach can save a great deal of work, and by means of its logical economy bring into sharper relief the decisive geometrical features of a given theory.
    
    In many cases, such as those of synthetic category theory, synthetic domain theory, synthetic differential geometry, and synthetic homotopy theory, it is possible to extend a given axiomatic model to a synthetic model~\cite{fiore-plotkin:1996,kapulkin-lumsdaine:2021,riehl-shulman:2017,shulman:2019,dubuc:1979,moerdijk-reyes:1991} and in some cases, this extension takes the form of a sharp representation theorem~\cite{fiore-rosolini:1997:cpos}.
    As these examples suggest, it is nearly always arranged in advance that results obtained using synthetic methods can be translated back to the original axiomatic models that are of interest; indeed, the goal of synthetic mathematics is not usually to set up a new ``foundation'' that competes with the old one, but rather to introduce new vocabulary that simplifies and clarifies the existing foundations. To put it another way, synthetic differential geometry does not ``compete with'' the theory of smooth manifolds any more than the complex plane ``competes with'' the real line.

  \end{xsect}

  \begin{xsect}{Synthetic domain theory}
    By the 1980s, classical domain theory had grown into a forbidding minefield of subtly different categories of domains that each presented a mix of advantages and disadvantages, without yet giving way to a simple enough general-purpose axiomatisation that could be used by non-specialists. Around this time, Dana Scott suggested a radical synthetic reformulation of domain theory in which domains would be special ``sets'' and all functions would be automatically continuous.
    
    Initial contributions to Scott's synthetic programme were carried out in the mid 1980s and early 1990s by Rosolini~\cite{rosolini:1986} and continued by several other authors including Freyd, Hyland, Mulry, Phoa, Scott, and Taylor~\cite{phoa:1991,hyland:1991,taylor:1991,freyd-mulry-rosolini-scott:1992}. During the 1990s, two notable directions emerged:
    \begin{enumerate}
      \item Fiore, Makkai, Plotkin, Power, and Rosolini~\cite{fiore-rosolini:1997:cpos,fiore-plotkin-power:1997,fiore-plotkin:1996,makkai-rosolini:1997} made great strides in understanding the formal relationship between axiomatic and synthetic domain theory, focussing on sheaf models of the latter.
      \item Reus and Streicher~\cite{reus-streicher:1993,reus-streicher:1999,reus-streicher:2012} initiated the study of \emph{General Synthetic Domain Theory}, which aims to study synthetic domains in type theory while avoiding non-elementary as well as order-theoretic assumptions.
    \end{enumerate}

    The adequacy of synthetic domains for denotational semantics of recursive types was studied by Simpson~\cite{simpson:2004}, who developed \emph{en passant} a very general version of synthetic domain theory that avoided many axioms that are easy to refute in models. Simpson's version is stated in terms of algebraic set theory~\cite{joyal-moerdijk:1995}, but it seems that only Martin-L\"of type theory with large eliminations for W-types was needed.

    In all these axiomatisations, the main ingredient was an \emph{interval}---usually a dominance in the sense of Rosolini~\cite{rosolini:1986}, \ie a subuniverse of the subobject classifier $\Prop$ closed under internal sums and the true proposition. The purpose of the interval is two-fold: to induce a form of directed homotopy (\cf the information order), and to parameterise the domains of partial functions. Then, predomains are isolated by imposing \emph{orthogonality} laws---in other words, a predomain is defined to be a type that is internally right-orthogonal (\ie local) to some small collection of maps. Of course, there are many possible such localisations and each gives rise to a different flavour of domain---some of which have full information orders, and others of which permit general recursion and little else.
  \end{xsect}

  \begin{xsect}[sec:synthetic-higher-category-theory]{Synthetic higher category theory}
    Inspired by the \emph{complete Segal space} model of higher categories~\cite{rezk:2001}, Riehl and Shulman~\cite{riehl-shulman:2017} developed a synthetic model of ($\infty$,1)-categories, hereafter `$\infty$-categories', in a simplicial version of homotopy type theory. The idea is that Martin-L\"of's identity types capture the \emph{invertible} higher structure of a given $\infty$-category, whereas the simplicial paths capture the \emph{directed} structure. The simplicial extension consists of a judgemental interval object and some logical operations that enable various geometrical shapes to be formed---including higher simplices and horns.

    Riehl and Shulman define synthetic $\infty$-categories by means of two orthogonality conditions: Segal completeness and Rezk completeness. Segal completeness allows directed arrows to be composed, and Rezk completeness ensures that invertible arrows coincide with identifications. The internal statement of Segal completeness was, in fact, already known by synthetic domain theorists in the 1990s, who used it to isolate a class of types that are preordered by their path relation~\cite{fiore-rosolini:1997:cpos}.
  \end{xsect}

  \begin{xsect}{Outline of this paper}

    Here we outline the main contributions of this paper and their relationship to other recent works in this area.
    
    \begin{xsect}{Synthesis of domains and higher categories}
      We take the first steps toward a unified foundation for synthetic domains and higher categories within Voevodsky's univalent foundations~\cite{hottbook}. Rather than imposing a number of axioms governing the interval at the outset as in all other papers in these two areas, we instead work with an \emph{arbitrary} interval object at all times and impose necessary assumptions only where they are used. This is needed in order to ensure the long-term applicability of our results in an environment where the axioms are constantly changing.

      \begin{xsect}{Comparison with Riehl and Shulman}
        We follow Gratzer, Weinberger and Buchholtz~\cite{gratzer-weinberger-buchholtz:2024} in eschewing the intricate simplicial judgemental structure introduced by Riehl and Shulman~\cite{riehl-shulman:2017} and simply assume an interval without any definitional laws or extension types. Although it might seem that the definitional laws are important to avoid getting buried under coherences, our experience suggests the opposite---at least in our case. Not only are abstract proofs based on universal properties equally easy without the definitional laws, we are constantly using various geometrical shapes (defined by pushouts in homotopy type theory) whose definitionally strict variants might not have been fibrant anyway.
      \end{xsect}

      \begin{xsect}{Comparison with Gratzer, Weinberger, and Buchholtz}
        In contrast to Gratzer~\etal~\cite{gratzer-weinberger-buchholtz:2024}, we avoid the non-type-theoretic assumption that the simplices form a global separator.\footnote{By ``non-type-theoretic'' we mean that such an assumption cannot be correctly stated as an axiom in type theory, because it applies \emph{only} to things defined in the empty context.} Aside from the generality (which we expect to be important for applications in domain theory), the immediate benefit of avoiding non-type-theoretic assumptions is that we can work in pure homotopy type theory without bringing in the syntactically challenging modal operators of Gratzer~\etal, around whom a consensus for rigorous \emph{informal} mathematics in the style of the HoTT Book~\cite{hottbook} has yet to emerge. Admittedly, without adding such modal operators we cannot express the duality involutions of synthetic categories---but we did not need these.
      \end{xsect}
    \end{xsect}

    \begin{xsect}{A new orthogonality law: \emph{based} Segal completeness}
      Synthetic $\infty$-categories are defined by Riehl and Shulman~\cite{riehl-shulman:2017} to be types satisfying the Segal and Rezk conditions, as we summarised in \S~\ref{sec:synthetic-higher-category-theory}. We have discovered a strengthening of the Segal condition that, when applied to a synthetic partial order (\ie a suitably truncated synthetic $\infty$-category), identifies the Sierpi\'nski cone with the partial map classifier. The ordinary Segal condition requires that a type $X$ be right orthogonal to the inner horn inclusion $\Horn\hookrightarrow\ICh{\Two}$; our new orthogonality law (\S~\ref{sec:based-segal-completeness}), dubbed \emph{based Segal completeness}, requires that $\II\times X$ be right orthogonal to $\Horn\hookrightarrow\ICh{\Two}$ in the slice over $\II$.
    \end{xsect}

    \begin{xsect}{Partial map classifiers for synthetic higher categories}
      En passant, we show that under appropriate assumptions on the interval, various notions of synthetic $\infty$-category are closed under partial map classifiers (Corollary~\ref{cor:infty-cat-pmc} via Theorems~\ref{thm:interval-is-based-segal-complete}, \ref{thm:interval-is-segal-complete}, \ref{thm:segal-complete-vs-well-complete}, and \ref{thm:based-segal-complete-vs-well-complete}).
    \end{xsect}

    \begin{xsect}{Comparing the Sierpi\'nski cone and the partial map classifier}
      Our main object of study is the comparison between the Sierpi\'nski cone (which freely adds an initial object to a synthetic category) and the partial map classifier. Although they do not coincide in general, we use our \emph{based} Segal condition to exhibit a reflective subuniverse in which they coincide (our main result, Corollary~\ref{cor:good-reflective-subuniverse} via Theorem~\ref{thm:main-result}); the upshot is that in such cases, the partial map classifier is the free cocompletion by an initial object and therefore has an additional \emph{monotone case analysis} principle that allows one to pattern match on whether a partial element is defined or not—as one often does in denotational semantics. 
      
      As we discuss in \S~\ref{sec:conclusion}, our result is not entirely satisfactory as it applies only to synthetic posets and not to general $\infty$-categories as we might have hoped. This does not, of course, pose any obstacle to applications in synthetic domain theory where the possibility of higher-dimensional domains is only beginning to emerge.
    \end{xsect}
  \end{xsect}
\end{xsect}
\begin{xsect}{Synthetic geometry of an interval}

  \begin{quote} 
    Everything that follows is written in the vernacular of univalent foundations; where the HoTT Book~\cite{hottbook} says “there merely exists” we shall simply say “there exists” in accordance with the traditions of conventional mathematics.
  \end{quote}

  Although synthetic category theory~\cite{riehl-shulman:2017} and synthetic domain theory~\cite{rosolini:1986,hyland:1991,phoa:1991} both build their axiomatics on an interval, they do so in different ways. In synthetic domain theory, one considers \emph{specifically} the interval objects that arise by restricting the ``standard'' (Lawvere) interval $\prn{\Prop,\bot,\top}$ to a dominance~\cite{rosolini:1986}, whereas no such assumptions are made in the synthetic theory of categories. As we aim to provide workhorse lemmas that can be used to advance and unify both synthetic domain and category theory, we will be especially careful to impose minimal assumptions on the interval.

  \begin{definition}
    We define an \emph{interval} simply to be a 01-bounded meet semilattice $\prn{\II,0,1,\sqcap}$;\footnote{So $\II$ is, in particular, a \emph{set} in the sense of univalent foundations~\cite{hottbook}.} we will write $i\sqsubseteq j$ for the induced partial order $i\sqcap j = i$.
  \end{definition}

  For an interval $\II$, we shall respectively write $\IsT{-},\IsF{-} \colon \II\to \Prop$ for the characteristic functions of the subsets $\brc{1},\brc{0}\subseteq\II$:
  \[ \IsT{i} :\equiv \prn{i=1},\qquad
     \IsF{i} :\equiv \prn{i=0}
  \]

  We have assumed very little about the interval, but we can nonetheless see straightaway that $\IsT{-}\colon\II\to\Prop$ preserves finite meets. Beyond this, one must consider more specific kinds of intervals to narrow down the behaviour of $\IsT{-}$.

  For example, when $\IsT{-}$ preserves the empty join $0$, we say that $\II$ is \emph{consistent} because it means that $0\not=1$. We have not asked for any other joins, but in many examples, $\II$ has \emph{stable binary joins} $i\sqcup j$ and thus forms a bounded distributive lattice. When these joins are preserved by $\IsT{-}$, we say that $\II$ has the \emph{disjunction property} because we would have $\IsT{i\sqcup j}\Leftrightarrow\IsT{i}\lor\IsT{j}$.

      \begin{lemma}
        Any totally ordered interval is a bounded distributive lattice satisfying the disjunction principle.
      \end{lemma}
  
      \begin{IEEEproof}
        Let $\II$ be totally ordered.
        Fixing $i,j:\II$, we define the join $i\sqcup j$ to be $j$ when $i\sqsubseteq j$ and $i$ when $j\sqsubseteq i$. This assignment defines a function by our assumption that $\II$ is totally ordered. Next we must check that $\IsT{i\sqcup j} = \IsT{i}\lor\IsT{j}$; this holds immediately in case either $i\sqsubseteq j$ or $j\sqsubseteq i$.
      \end{IEEEproof}

\begin{xsect}{Conservative intervals and open embeddings}
  It is also reasonable to compare $\II$ to its image in $\Prop$. %
  \begin{definition}
    An interval $\II$ is called \emph{conservative} when either of the following equivalent properties hold:
    \begin{enumerate}
      \item $\IsT{-}\colon \II\to \Prop$ is an embedding of sets, so we have $\IsT{i}=\IsT{j}$ if and only if $i=j$.
      \item $\IsT{-}\colon\II\to\Prop$ is an order-embedding, so we have $\IsT{i}\to\IsT{j}$ if and only if $i\sqsubseteq j$.
    \end{enumerate}
  \end{definition}

  \begin{IEEEproof}
    The second condition clearly implies the first. 
    On the other hand, we fix $i,j:\II$ such that $\IsT{i}\to\IsT{j}$ to check that $i\sqsubseteq j$. In other words, we must check that if $\IsT{i}\land\IsT{j}=\IsT{i}$ holds then we have $i\sqcap j = i$. Because $\IsT{-}$ is preserves finite meets, we have $\IsT{i\sqcap j} = \IsT{i}$ and thus $i\sqcap j=i$.
  \end{IEEEproof}

  \begin{definition}
    We shall call an embedding $A\hookrightarrow{B}$  \emph{open} when its characteristic function $B\to\Prop$ factors through $\IsT{-}\colon\II\to\Prop$. We shall write $\OpenEmbeddings$ for the class of all open embeddings.
  \end{definition}

  In the case of a conservative interval $\II$, we can by definition identify the partial order $\II$ with its image $\OpenProp\subseteq\Prop$. When $\II$ is \emph{consistent}, it follows that the entire 01-bounded meet semilattice structure $\prn{\II,0,1,\sqcap}$ can be identified with the structure $\prn{\OpenProp,\bot,\top,\land}$ on the image inherited from $\Prop$. Likewise, when $\II$ has stable joins satisfying the disjunction property, the bounded distributive lattice $\prn{\II,0,1,\sqcap,\sqcup}$ can be identified with the structure $\prn{\OpenProp,\bot,\top,\land,\lor}$.
\end{xsect}
\begin{xsect}{The interval as a universe}
  We can think of the map $\IsT{-}\colon\II\to\Prop$ as exhibiting $\II$ as a \emph{universe} of propositions of the form $\IsT{i}$ for a given $i:\II$. When $\II$ is conservative and so $\IsT{-}$ is an embedding, then $\II=\OpenProp$ is a \emph{univalent universe} of propositions.
  
  Because finite meets are preserved by $\IsT{-}$ automatically, we can view the meet structure on $\II$ as \emph{closing} this universe of propositions under products. When $\II$ is a bounded distributive lattice satisfying the disjunction principle, this the same as to say that the corresponding universe of propositions is closed under disjunctions. 
  
  \begin{xsect}{Closure under internal sums}
    A further condition that can be stated most sensibly in terms of this universe is closure under \emph{internal sums}, \ie Martin-L\"of's $\Sigma$-types. Awodey~\cite{awodey:2018:natural-models} explains very clearly that this structure can be expressed in terms of the polynomial type constructor $L\prn{X} = \sum_{\prn{i:\II}}X\Sup{\IsT{i}}$. In particular, we instantiate Awodey's definition to the interval as follows:

    \begin{definition}\label{def:internal-sums} 
      An \emph{internal sum} structure on the interval is defined to be a cartesian morphism of polynomials $L\circ L\to L$. More explicitly, this is simply the following pullback square:
      \[ 
        \begin{tikzcd}
          \brc{\prn{1,\Lam{\_}1}}
            \arrow[r]
            \arrow[d,hookrightarrow] 
            \arrow[dr,phantom,very near start,"\lrcorner"]
          & 
          \brc{1}
            \arrow[d,hookrightarrow]\\
          L\prn{\II}
            \arrow[r,dashed,swap,"\sum"]
          & 
          \II
        \end{tikzcd}
      \]

      An internal sum structure is evidently a special kind of \emph{algebra} for the polynomial endofunctor $L$ that ultimately turns the latter into a certain kind of polynomial pseudomonad—a perspective developed in more detail and generality by Awodey and Newstead~\cite{awodey-newstead:2018}.
    \end{definition}

    When $\II$ is conservative, the internal sum structure is unique and exhibits $L$ as a (strict) polynomial monad in the sense of Gambino and Kock~\cite{gambino-kock:2013}. One of our goals, however, is to limit the assumptions we place on $\II$ as much as possible whilst setting up foundations. For example, the following condition is automatic for a conservative interval equipped with an internal sum structure, but is decisive for many important lemmas in case we have not assumed conservativity:

    \begin{definition}\label{def:factoring-meets}
      An internal sum structure $\sum$ on an interval $\II$ is said to \emph{factor binary meets} when the following commutes:
      \[ 
        \begin{tikzcd}
          \II\times \II
            \arrow[dr,swap,"\prn{i,j}\mapsto \prn{i,\Lam{\_}j}"]
            \arrow[rr,"\sqcap"]
          &&
          \II
          \\
          &
          L\prn{\II}
          \arrow[ur,swap,"\sum"]
        \end{tikzcd}
      \] 
    \end{definition}
  \end{xsect}

  \begin{xsect}{Rosolini's dominances}
    Synthetic domain theory usually starts from a \emph{dominance}~\cite{rosolini:1986}, which is a subuniverse of $\Prop$ closed under $\top$ and internal sums in the sense of Definition~\ref{def:internal-sums}. It is indeed true that a \emph{conservative} interval closed under internal sums in our sense is a dominance, but we do not start directly from here as we are interested in keeping track of where exactly various assumptions are used.
  \end{xsect}
\end{xsect}
\begin{xsect}[sec:geometry]{Representing geometrical figures using the interval}
  One of the roles of an interval $\II$ is to parameterise a (potentially directed) notion of \emph{path} or \emph{homotopy}. In particular, a path in a type $X$ from $x$ to $y$ can be defined to be a function $f\colon \II\to X$ equipped with identifications $f(0)=x$ and $f(1) = y$; we may write $f\colon x \rightsquigarrow y$ for such a path. A priori, the path structure of an arbitrary type $X$ behaves like a directed pseudograph: for any two vertices of $X$ there is a type of edges between them, but we have no further operations for composing, inverting, or identifying these edges.

  It will be useful to isolate classes of types $X$ for which the induced notion of homotopy behaves in more specific ways. In order to do this, we must be able to refer to various figures drawn in $X$ such as \emph{simplices}, \emph{inner horns}, and \emph{biinvertible arrows}. All of these figures can be defined \emph{representably} in terms of functions $S\to X$ where $S$ is a generic ``figure shape'' built up from the interval.

  \begin{xsect}[sec:simplices]{Representing simplices}

    Keeping in mind that paths in arbitrary types $X$ need not be composable in any sense, we must define what it means for a given path $x\rightsquigarrow z$ to be the composite of some chain $x\rightsquigarrow y \rightsquigarrow z$ and so on. The usual way to do this is by means of some kind of \emph{simplicial} structure, which is handily supplied by the interval $\II$ itself.

    For finite $n$, we define $\ICh{n}$ to be the subposet of the $n$-cube $\II^n$ spanned by \emph{descending} sequences $\prn{i_1\sqsupseteq \ldots\sqsupseteq i_n}$. In particular, we have $\ICh{\Empty}=\mathbf{1}$ and $\ICh{\One} = \II$ and $\ICh{\Two} = \Compr{\prn{i,j}}{i\sqsupseteq j}$ and so on. We shall refer to $\ICh{n}$ as the \emph{$n$-simplex}. 

    We can probe a given type $X$ to discover a variety of simplicial figures:
    \begin{enumerate}
      \item A map $\alpha\colon \ICh{\Empty}\to X$ determines a single point $\alpha(*)\colon X$.
      \item A map $\alpha\colon \ICh{\One}\to X$ determines a path $\alpha\prn{0} \rightsquigarrow \alpha\prn{1}$.
      \item A map $\alpha\colon \ICh{\Two}\to X$ determines a (filled) \emph{triangle} in $X$:
        \[ 
          \begin{tikzcd}[column sep=large]
            \alpha\prn{0\sqsupseteq 0} 
            \arrow[r,rightsquigarrow, "j\mathpunct{.}\alpha\prn{j\sqsupseteq 0}"] 
            \arrow[dr,rightsquigarrow,swap,sloped,"k\mathpunct{.}\alpha\prn{k\sqsupseteq k}"]
            & \alpha\prn{1\sqsupseteq0}
            \arrow[d,rightsquigarrow,"i\mathpunct{.}\alpha\prn{1\sqsupseteq i}"]
            \\ 
            & \alpha\prn{1\sqsupseteq1}
          \end{tikzcd}
        \]  
        and so on.
    \end{enumerate}
  \end{xsect}

  \begin{xsect}[sec:inner-horns]{Representing inner horns}
    An \emph{inner horn} in a type $X$ is defined to be a pair of compatible paths $x_0\rightsquigarrow x_1 \rightsquigarrow x_2$. A figure like is represented by the ``walking'' inner horn $\Horn$ which we can describe by glueing the left endpoint of one copy of the interval onto the right endpoint of another copy of the interval:
    \[ 
      \begin{tikzcd}
        \brc{0}\times\brc{1} \arrow[r,hookrightarrow,"1"] \arrow[d,hookrightarrow,swap,"0"] \arrow[dr,very near end,phantom,"\ulcorner"] & \II\arrow[d,hookrightarrow,dashed]\\ 
        \II\arrow[r,hookrightarrow,dashed] & \Horn
      \end{tikzcd}
    \]

    The following square determines an embedding $\Horn\hookrightarrow \ICh{\Two}$ by the universal property of $\Horn$:
    \[ 
      \begin{tikzcd}
        \brc{0}\times\brc{1} \arrow[r,hookrightarrow,"1"] \arrow[d,hookrightarrow,swap,"0"] & \II\arrow[d,hookrightarrow,"\prn{-\sqsupseteq 0}"]\\ 
        \II\arrow[r,hookrightarrow,swap,"\prn{1\sqsupseteq -}"] & \ICh{\Two}
      \end{tikzcd}
    \] 

    Inspecting the fibres of the induced embedding $\Horn\hookrightarrow\ICh{\Two}$ reveals the alternative type theoretic formulation of the walking inner horn used by Riehl and Shulman~\cite{riehl-shulman:2017}, which we can use as a type theoretically convenient definition:
    \[ 
      \Horn :\equiv \Compr{\prn{i\sqsupseteq j}\in \ICh{\Two}}{\IsF{j}\lor \IsT{i}}
    \] 
  \end{xsect}

  \begin{xsect}[sec:equivalences]{Representing equivalences}
    An \emph{equivalence} in a type $X$ is defined to be a path $f\colon x\rightsquigarrow y$ together with a section $s\colon y\rightsquigarrow x$ and a retraction $r\colon y\rightsquigarrow x$. Because we have not assumed that paths can be composed, the witnesses that $s$ and $r$ are sections and retractions of $f$ respectively must be witnessed by \emph{triangles}, as represented by $\ICh{\Two}$. 
    
    All the data of such a complicated figure can be represented by the \emph{walking equivalence} $\WalkingBiinv$, whose construction we recall from Buchholtz and Weinberger~\cite{buchholtz-weinberger:2023}. In particular, we define $\WalkingBiinv$ to be the colimit of the following diagram:
    \[
      \begin{tikzcd}[cramped, column sep=small]
        & 
        \ICh{\One}
          \arrow[dl,sloped,"\prn{}"]
          \arrow[dr,sloped,hookrightarrow,"\prn{-\sqsupseteq-}"] 
        && 
        \ICh{\One}
          \arrow[dl,hookrightarrow,sloped,"\prn{-\sqsupseteq0}"]
          \arrow[dr,hookrightarrow,sloped,"\prn{1\sqsupseteq-}"] 
        && 
          \ICh{\One}
            \arrow[dl,sloped,hookrightarrow,"\prn{-\sqsupseteq-}"]
            \arrow[dr,sloped,"\prn{}"]
        \\
        \ICh{\Empty} && \ICh{\Two} && \ICh{\Two} && \ICh{\Empty}
      \end{tikzcd}
    \] 
  \end{xsect}

  \begin{remark}
    To understand the definition of the walking equivalence $\WalkingBiinv$, visualise a cocone under the diagram:
    \[
      \begin{tikzcd}[cramped, column sep=small]
        & 
          \ICh{\One}
            \arrow[dl,sloped,"\prn{}"]
            \arrow[dr,sloped,hookrightarrow,"\prn{-\sqsupseteq-}"]
            \arrow[dddrr,bend right=30,sloped,"\mathsf{id}_{x}" description] 
        && 
          \ICh{\One}
            \arrow[dl,sloped,hookrightarrow,"\prn{-\sqsupseteq 0}"]
            \arrow[dr,sloped,hookrightarrow,"\prn{1\sqsupseteq0}"]
            \arrow[ddd,"f" description] 
        && 
          \ICh{\One}
            \arrow[dl,sloped,hookrightarrow,"\prn{-\sqsupseteq-}"]
            \arrow[dr,sloped,"\prn{}"]
            \arrow[dddll,bend left=30,sloped,"\mathsf{id}_y" description]
        \\
        \ICh{\Empty}
          \arrow[ddrrr,bend right=30,sloped,"x" description] 
        &&
        \ICh{\Two}
          \arrow[ddr,"\rho" description] 
        && 
          \ICh{\Two}
          \arrow[ddl,"\sigma" description] 
        && 
          \ICh{\Empty}
            \arrow[ddlll,bend left=30,"y" description]\\\\
        &&&
          A
      \end{tikzcd}
    \] 
    
    Recalling the geometrical interpretation of $\ICh{n}$-figures in $A$ as simplices, such a cocone gives precisely a pair of 0-cells $x,y:A$, a 1-cell $f\colon x\to y$, a retraction figure $\rho\colon r_f\circ f = 1_x$ and a section figure $\sigma\colon f\circ s_f = 1_y$, where the 1-cells $s_f$ and $r_f$ are obtained by restricting $\rho$ and $\sigma$ along $\prn{1\sqsupseteq -}\colon \ICh{\One}\hookrightarrow \ICh{\Two}$ and $\prn{-\sqsupseteq 0}\colon \ICh{\One}\hookrightarrow \ICh{\Two}$ respectively. When $A$ is Segal complete in the sense that we shall define in \S~\ref{sec:segal-completness}, the data above defines an actual equivalence in $A$.
  \end{remark}
\end{xsect}
\begin{xsect}{Constructions involving the interval}
  In this section we introduce the two constructions that this paper aims to relate: the weak partial map classifier $L\prn{X}$  and the Sierpi\'nski cone construction $X_\bot$.

  \begin{xsect}{The partial map classifier}
    Any family $B \colon A\to \Type$ determines a \emph{polynomial} type constructor $X\mapsto \sum\Sub{\prn{a:A}}X^{B\prn{a}}$.  In the case of $\IsT{-}\colon \II\to\Prop\hookrightarrow\Type$, the polynomial $L(X) :\equiv \sum\Sub{\prn{i:\II}}X\Sup{\IsT{i}}$ forms the base for the \emph{partial map classifier} $\eta_X\colon X \hookrightarrow L\prn{X}$ that sends $x$ to the pair $\prn{1,\Lam{\_}x}$.
    
    The universal property of $\eta_X\colon X\hookrightarrow L\prn{X}$ is that of the \emph{partial product} with $\brc{1}\hookrightarrow\II$: in other words, the function space $Y\to L\prn{X}$ is canonically isomorphic to the type of ``partial maps'' from $Y$ into $X$ with supports valued in $\II$:
    \[ \mathsf{PMap}(Y,X) :\equiv \textstyle\sum\Sub{\prn{s\colon Y\to\II}} \Compr{y:Y}{\IsT{s\prn{y}}}\to X\]
    
    As we have not assumed that $\IsT{-}\colon \II\to \Prop$ is an embedding, our notion of ``partial map'' is naturally a bit intensional. If it is an embedding, then a partial map from $Y$ to $X$ is uniquely determined by an open subspace of $Y$ with a map to $X$.
  \end{xsect}

  \begin{xsect}{The Sierpi\'nski cone}
    The Sierpi\'nski cone of a type $X$ is a 2-dimensional colimit, which can be equivalently described by a conical colimit using the interval $\II$. 

    \begin{definition}
      The \emph{Sierpi\'nski cone} $\brc{\bot}\hookrightarrow X_\bot\hookleftarrow X$ of a type $X$ is defined to be the following co-comma object, which we equivalently describe as a pushout:
      \[ 
        \begin{tikzcd}
          X\arrow[r,equal]\arrow[d,->] & X\arrow[d, hookrightarrow]\\ 
          \brc{\bot}\arrow[r,hookrightarrow]\arrow[ur,phantom,"{\Uparrow}\varsigma_X"] & X_\bot
        \end{tikzcd}
        \qquad
        \begin{tikzcd}
          X\arrow[r,"0\times X"]\arrow[d]\arrow[dr,very near end,phantom,"\ulcorner"] & \II\times X\arrow[d] & X\arrow[l,hookrightarrow,swap,"1\times X"]\arrow[dl,hookrightarrow]\\ 
          \brc{\bot}\arrow[r,hookrightarrow] & X_\bot
        \end{tikzcd}
      \]
    \end{definition}

\begin{xsect}[sec:sierpinski-data]{Sierpi\'nski data and its structure identity principle}

  In this section, we characterise the identity types of function spaces of the form $C\Sup{X_\bot}$ without any truncation-level assumptions. First of all, the universal property of the pushout description of $X_\bot$ ensures that $C\Sup{X_\bot}$ is the following pullback:
  \[ 
    \begin{tikzcd}
      C^{X_\bot}
        \arrow[dr,phantom,very near start,"\lrcorner"]
        \arrow[r,"C^{\varsigma_X}"]
        \arrow[d,swap,"C^\bot"]
      &
      C^{\II\times X}
        \arrow[d,"C^{0\times X}"]
      \\ 
      C\Sup{\brc{\bot}}
        \arrow[r,swap,"C^!"]
      &
      C^X
    \end{tikzcd}
  \] 
  
  We therefore have a canonical isomorphism $C^{X_\bot}\cong\mathsf{SierpData}_X\prn{C}$ with the latter defined like so:
  \begin{align*}
    &\mathsf{SierpData}_X\prn{C} :\equiv
    \\
    &\quad
    \textstyle
    \sum_{\prn{c^\bot: C}}
    \sum_{\prn{c^\varsigma: \II\times X\to C}}
    \prod_{\prn{x:X}} c^\bot = c^\varsigma\prn{0,x}
  \end{align*}

  Next, we characterise the identity types $f = g$ for $f,g: \mathsf{SierpData}_X\prn{C}$. In particular, we define the following reflexive graph structure $\prn{\approx, \mathsf{rx}}$ on $\mathsf{SierpData}_X\prn{C}$, where $\bullet$ composes identifications:
  \begin{align*}
    &\prn{f^\bot,f^\varsigma,H_f}\approx \prn{g^\bot,g^\varsigma, H_g} :\equiv
    \\ 
    &\quad 
    \textstyle
    \sum\Sub{
      \prn{\alpha_\bot\colon f^\bot = g^\bot}
    }
    \sum\Sub{
      \prn{\alpha_\varsigma\colon f^\varsigma \sim g^\varsigma}
    }
    \\ 
    &\quad
    \textstyle
    \prod\Sub{
      \prn{x:X}
    }
    H_f\prn{x} \bullet \alpha_\varsigma\prn{0,x} = \alpha_\bot \bullet H_g\prn{x}
    \\ 
    &\mathsf{rx}\Sub{\prn{f^\bot,f^\varsigma,H_f}}
    :\equiv
    \\
    &\quad 
    \begin{pmatrix*}[l]
      \mathsf{refl}_{f^\bot},\\
      \Lam{\prn{i,x}}\mathsf{refl}_{f^\varsigma\prn{i,x}},\\
      \Lam{x}
      \mathsf{runit}_{H_f\prn{x}} \bullet 
      \mathsf{lunit}_{H_f\prn{x}}^{-1}
    \end{pmatrix*}
  \end{align*}

  \begin{lemma}[Structure identity principle for Sierpi\'nski data]\label{lem:sierpinski-data-sip}
    The reflexive graph $\prn{\mathsf{SierpData}_X\prn{C},\approx, \mathsf{rx}}$ is univalent in the sense of Schipp von Branitz and Buchholtz~\cite{schipp-von-branitz-buchholtz:2021,schipp-von-branitz:2020:thesis}: for each $f:\mathsf{SierpData}_X\prn{C}$, the based edge space $\sum\Sub{\prn{g:\mathsf{SierpData}_X\prn{C}}}f\approx g$ is contractible.
  \end{lemma}

  In other words, the canonical map $f=g\to f\approx g$ determined by the reflexivity datum is an isomorphism.
\end{xsect}
 
  \end{xsect}

\end{xsect} 
\end{xsect} %

\begin{xsect}{Sierpi\'nski cone vs.\ partial map classifier}
  In many concrete 2-categories of spaces (like cpos, categories, \etc), the Sierpi\'nski cone is in fact the partial map classifier determined by the interval, \ie we have $X_\bot \cong L\prn{X}$. This will \emph{not} be the case for many $X$ in the synthetic setting because the inclusions of various synthetic notions of space into all types will not preserve many colimits. We can, however, investigate the relationship between these two constructions by means of a canonical comparison map from the Sierpi\'nski cone.

  \begin{xsect}[sec:scone-type-theoretic]{Type theoretic computation of the Sierpi\'nski cone}

    An equivalent type theoretical description of the Sierpi\'nski cone clarifies the comparison with the partial map classifier.\footnote{We are thankful to Reid Barton for pointing this out.} In particular, the following is seen to be a co-comma square, using the fact that $\sum_{\II}:\mathbf{Type}/\II\to \mathbf{Type}$ preserves colimits:\footnote{Here $X*Y$ denotes the \emph{join} of types, \ie the pushout of the product span $X\gets X\times Y\to Y$. When $X$ and $Y$ are propositions, this is just disjunction.}
    \[ 
      \begin{tikzcd}[column sep=huge]
        X\arrow[r,equal]\arrow[d,->] & X\arrow[d, hookrightarrow, "\prn{1, \mathsf{inr}}"]\\ 
        \brc{\prn{0,\mathsf{inl}(*)}}\arrow[r,hookrightarrow]\arrow[ur,phantom,"\mathllap{{\Uparrow}\Lam{i}}\prn{i,\mathsf{glue}_{\IsF{i}*X}}"] & \sum_{\prn{i:\II}}\IsF{i}*X
      \end{tikzcd}
    \] 

    \begin{IEEEproof}
      We will work in the slice over a generic element $i:\II$. The join $\IsF{i}*X$ is the following pushout:
      \[ 
        \begin{tikzcd}
          \IsF{i}\times X 
            \arrow[r]
            \arrow[d]
            \arrow[dr,phantom,very near end,"\ulcorner"]
          & 
          X 
            \arrow[d]
          \\ 
          \IsF{i} 
            \arrow[r]
          & 
          \IsF{i}*X
        \end{tikzcd}
      \]

      The dependent sum functor $\sum_{\prn{i:\II}}$ is left adjoint to weakening along $i:\II$, and so it preserves colimits. Therefore, the following is a pushout diagram:
      \[ 
        \begin{tikzcd}
          \sum_{\prn{i:\II}}{\IsF{i}\times X }
            \arrow[r]
            \arrow[d]
            \arrow[dr,phantom,very near end,"\ulcorner"]
          & 
          \II\times X 
            \arrow[d]
          \\ 
          \sum_{\prn{i:\II}}\IsF{i} 
            \arrow[r]
          & 
          \sum_{\prn{i:\II}}\IsF{i}*X
        \end{tikzcd}
      \]

      We can simplify the left-hand vertices because they constrain their parameter entirely:
      \[ 
        \begin{tikzcd}
          \brc{0}\times X
            \arrow[r]
            \arrow[d]
            \arrow[dr,phantom,very near end,"\ulcorner"]
          & 
          \II\times X 
            \arrow[d]
          \\ 
          \brc{0}
            \arrow[r]
          & 
          \sum_{\prn{i:\II}}\IsF{i}*X
        \end{tikzcd}
      \]

      This is precisely the description of the Sierpi\'nski cone of $X$ as a conical colimit.
    \end{IEEEproof}

    Therefore, we can define $X_\bot :\equiv \sum_{\prn{i:\II}}\IsF{i}*X$.

  \end{xsect}

  \begin{xsect}{Conceptual comparison of \texorpdfstring{$L(X)$}{L(X)} with \texorpdfstring{$X_\bot$}{Scone(X)}}
    The type theoretic presentation that we have given in \S~\ref{sec:scone-type-theoretic} provokes a direct comparison to the partial map classifier. Recall that for a proposition $P$, the function space $X^P$ is the \emph{open modality} associated to $P$ and the join $P*X$ is the \emph{closed modality} associated to $P$. By definition, $L\prn{X}$ classifies partial maps defined within an open subspace; our analysis suggests that the Sierpi\'nski cone ought to be thought of as classifying some kind of (quasi?-)partial map defined ``away from'' the complement of an open subspace.
    
    The partial map classifier and the Sierpi\'nski cone are therefore two vertices in a matrix (see Table~\ref{table:domain-and-modality-matrix}) of different ways to express mappings that are not completely defined: 
    \begin{enumerate}
      \item The \emph{domain} of definedness can be described in an open or a closed way.
      \item The \emph{nature} of definedness itself can be expressed via an open modality or a closed modality.
    \end{enumerate}
    Thus the partial map classifier describes open definedness on an open domain; in contrast, the Sierpi\'nski cone describes \emph{closed} definedness on a \emph{closed} domain. Hyland's co-partial map classifier~\cite{hyland:1991} describes open definedness on a closed domain, and the final vertex (closed definedness on an open domain) is a dual shape to the Sierpi\'nski cone that freely adjoins a terminal element rather than an initial element.

    \begin{table*}
      \begin{center}
        \begin{tabular}{l|ll}
          \toprule 
          &\textbf{Open Modality} & \textbf{Closed Modality}
          \\
          \midrule
          \textbf{Open Domain}
          &
          Partial Map Classifier: $L\prn{X} = \sum\Sub{\prn{i:\II}}X\Sup{\IsT{i}}$
          &
          Sierpi\'nski Co-cone: $X_\top = \sum\Sub{\prn{i:\II}}\IsT{i}*X$
          \\ 
          \textbf{Closed Domain}
          &
          Co-Partial Map Classifier: $T\prn{X} = \sum\Sub{\prn{i:\II}}X\Sup{\IsF{i}}$
          &
          Sierpi\'nski Cone: $X_\bot = \sum\Sub{\prn{i:\II}}\IsF{i}*X$
          \\
          \bottomrule
        \end{tabular}
      \end{center}
      \caption{Comparison matrix of (co-)partial map classifiers and Sierpi\'nski (co-)cones.}
      \label{table:domain-and-modality-matrix}
    \end{table*}
  \end{xsect}

  \begin{xsect}{Explicit comparison of the \texorpdfstring{$L(X)$}{L(X)} with \texorpdfstring{$X_\bot$}{Scone(X)}}
    \begin{lemma}[Comparison map]
      When $\II$ is consistent, there is an evident comparison map $\sigma_X\colon X_\bot\to L\prn{X}$ over $\II$ determined fibrewise by the universal arrow $\sigma_{X,i}\colon \IsF{i}*X\to X^{\IsT{i}}$ given by the following square:
      \[ 
        \begin{tikzcd}
          \IsF{i}\times X\arrow[r]\arrow[d] & X\arrow[d,"\mathsf{const}"] \\ 
          \IsF{i}\arrow[r,swap,"\mathsf{abort}"] & X^{\IsT{i}}
        \end{tikzcd}
      \] 

      The lower map is defined only by virtue of our assumption that $\II$ is consistent, and so $\IsF{i}\land\IsT{i} = \bot$.
    \end{lemma}

    \begin{remark}\label{rem:little-comparison-map}
      Note that when $X$ is of the form $\IsT{j}$, the comparison map $\sigma_{X}=\sigma_{\IsT{j}}\colon \IsT{j}_\bot\to L\IsT{j}$ exists \emph{regardless} of whether $\II$ is consistent, as the component $\IsF{i}\to \IsT{j}^{\IsT{i}}$ exists automatically because we always have $0\sqsubseteq j$.
    \end{remark}
    \begin{IEEEproof}
      In particular, the arrow $\IsF{i}\to \IsT{j}^{\IsT{i}}$ exists (necessarily uniquely) regardless of whether $\II$ is consistent:
      \begin{prooftree*}
        \infer0{
          j:\II\mid \cdot \vdash 0\leq j
        }
        \infer[double]1{
          i,j:\II \mid \IsF{i}\vdash i\leq j
        }
        \infer[double]1{
          i,j:\II \mid \IsF{i}\vdash \IsT{i}\Rightarrow \IsT{j}
        }
        \infer[double]1{
          i,j:\II\mid \IsF{i}\vdash \IsT{i}\Rightarrow\IsT{j}
        }
      \end{prooftree*}
      We have used only the fact that $0$ is the least element of $\II$.
    \end{IEEEproof}
  \end{xsect}

  \begin{xsect}[sec:display-of-inner-horn-inclusion]{Display of the inner horn inclusion}
    The incidence relation $\Horn\hookrightarrow\ICh{\Two}$ between the inner horn $\Horn$ and the walking triangle $\ICh{\Two}$ can be displayed over the interval by projecting the \emph{upper} endpoints as follows:
    \[ 
      \begin{tikzcd}
        \Horn\arrow[dr,sloped,swap,"\sqcup"]\arrow[rr,hookrightarrow] && \ICh{\Two}\arrow[dl,sloped,swap,"\sqcup"]
        \\ 
        &
        \II
      \end{tikzcd}
    \] 

    \begin{lemma}\label{lem:fibres-of-the-triangle}
      The walking triangle $\ICh{\Two}$ is the sum $\sum_{\prn{j:\II}}\II/j$ of all the slices $\II/j$ for $j:\II$, as witnessed by the following cartesian squares:
      \[ 
        \begin{tikzcd}
            \II/j
            \arrow[rr, hookrightarrow,"\prn{i\mid j\sqsupseteq i}\mapsto \prn{j\sqsupseteq i}"]
            \arrow[d]
            \arrow[dr,phantom,very near start,"\lrcorner"]
          &&
          \ICh{\Two}
          \arrow[d,"\sqcup"]
          \\ 
          \brc{j}
            \arrow[rr,hookrightarrow]
          &{}&
          \II 
        \end{tikzcd}
      \]
    \end{lemma}
    \begin{IEEEproof}
      For $\prn{j\sqsupseteq i}\in\ICh{\Two}$ we have $i\sqcup j = j$. Therefore, the fibre of ${\sqcup}\colon \ICh{\Two}\to\II$ over $j:\II$ is $\Compr{i:\II}{i\leq j} = \II/j$.
    \end{IEEEproof}

    \begin{lemma}\label{lem:fibres-of-the-horn}
      The walking inner horn $\Horn$ is the sum $\sum_{\prn{j:\II}}\IsT{j}_\bot$ of all the ``little'' Sierpi\'nski cones $\IsT{j}_\bot$ for $j:\II$, as each of the following squares is cartesian:
      \[ 
        \begin{tikzcd}[column sep=large]
          \IsT{j}_\bot
            \arrow[rr, hookrightarrow, "\prn{i\mid \IsF{i}\lor \IsT{j}}\mapsto \prn{j\sqsupseteq i}"]
            \arrow[d]
            \arrow[dr,phantom,very near start,"\lrcorner"]
          &&
          \Horn
          \arrow[d,"\sqcup"]
          \\ 
          \brc{j}
            \arrow[rr,hookrightarrow]
          &{} &
          \II 
        \end{tikzcd}
      \]
    \end{lemma}
    \begin{IEEEproof}
      We have $\Horn = \Compr{\prn{j,i}:\ICh{\Two}}{\IsF{i}\lor\IsT{j}}$ by definition; for $\prn{i\sqsubseteq j}\in \Horn$, we have $i\sqcup j = j$. Therefore, the fibre of ${\sqcup}\colon \Horn\to\II$ over $j:\II$ is $\Compr{i:\II}{\IsF{i}\lor\IsT{j}}\cong \IsT{j}_\bot$.
    \end{IEEEproof}

    \begin{corollary}
      The inclusion $\Horn\hookrightarrow\ICh{\Two}$ is the sum of all the inclusions $\IsT{j}_\bot\hookrightarrow \II/j$ for $j:\II$, as witnessed by the following cartesian squares:
      \[ 
        \begin{tikzcd}
          \IsT{j}_\bot
            \arrow[r,hookrightarrow]
            \arrow[d,hookrightarrow]
            \arrow[dr,phantom,very near start,"\lrcorner"]
          &
          \Horn
            \arrow[d,hookrightarrow]
          \\ 
          \II/j
            \arrow[r,hookrightarrow]
            \arrow[d]
            \arrow[dr,phantom,very near start,"\lrcorner"]
          &
          \ICh{\Two}
            \arrow[d,"\sqcup"]
          \\ 
          \brc{j}
            \arrow[r,hookrightarrow]
          &
          \II
        \end{tikzcd}
      \] 
    \end{corollary}

    \begin{remark}
      When $\II$ is conservative, the evident function  $\II/j\to L\IsT{j}$ over $\II$ is invertible as we have $i\sqsubseteq j \Leftrightarrow \IsT{j}^{\IsT{i}}$; in this case, therefore, the fibre of $\ICh{\Two}$ over $j:\II$ would be the ``little'' partial map classifier $L\IsT{j}$. Recalling Remark~\ref{rem:little-comparison-map}, this isomorphism then identifies the inclusion $\IsT{j}_\bot\hookrightarrow \II/j$ with the comparison map $\sigma_{\IsT{j}}\colon \IsT{j}_\bot\to L\IsT{j}$.
    \end{remark}

  \end{xsect}

  \begin{xsect}{The Sierpi\'nski cone is not the partial map classifier}
    It is now possible to see why there could not be many $X$ for which the comparison map $\sigma_X\colon X_\bot\to L\prn{X}$ is an isomorphism: if $X=\IsT{j}$, this would entail $j\in\brc{0,1}$!

    \begin{observation}\label{obs:scone-is-not-pmc}
      Let $\II$ be a consistent interval; if each ``little'' comparison map $\sigma_{\IsT{j}}\colon \IsT{j}_\bot\to L\IsT{j}$ is an isomorphism for $j:\II$, then $\prn{0,1}\colon \mathbf{2}\to\II$ is an isomorphism.
    \end{observation}
    \begin{IEEEproof}
      The fibres of $\pi\colon \IsT{j}_\bot\to\II$ and $\pi\colon L\IsT{j}\to \II$ over $j:\II$ are $\IsF{j}\lor\IsT{j}$ and $\IsT{j}^{\IsT{j}}$ respectively, which are only equivalent should $j$ be either $0$ or $1$.
    \end{IEEEproof}
  \end{xsect}
\end{xsect}
\begin{xsect}[sec:orthogonal-classes]{Local classes of types}
  We can now proceed to identify various local subuniverses that correspond to different kinds of synthetic space that play a role in synthetic (higher) category theory and domain theory.

  \begin{definition}\label{def:S-epi-iso}
    Let $\mathcal{R}$ be a class of types.
    \begin{enumerate}
      \item A function $f\colon A\to B$ is called an \emph{$\mathcal{R}$-epimorphism} when for any $R\in\mathcal{R}$ the map $R^f\colon R^B\to R^A$ induced by precomposition with $f\colon A\to B$ is an embedding. 
      \item A function $f\colon A\to B$ is called an \emph{$\mathcal{R}$-isomorphism} when for any $R\in \mathcal{R}$ the map $R^f\colon R^B\to R^A$ is an isomorphism.
    \end{enumerate}

    When $S$ is a single type, we will often write $S$ to denote the class $\mathcal{S}=\brc{S}$ in the context of this definition. 
  \end{definition}

  \begin{definition}
    Let $\mathcal{L}$ be a class of maps. A type $X$ is called \emph{$\mathcal{L}$-local} when every $f\colon A\to B\in \mathcal{L}$ is an $X$-isomorphism, \ie the restriction map $X^f\colon X^B\to X^A$ is an isomorphism.
  \end{definition}

  Given classes of maps $\mathcal{S}$ and $\mathcal{L}$, we define $\mathcal{S}^*\mathcal{L}$ to be the class of maps obtained by pulling back an element of $\mathcal{L}$ along an element of $\mathcal{S}$.

  \begin{table*}[t]
    \begin{center}
      \begin{tabular}{llll}
        \toprule
        \textbf{Property} & \textbf{Local With Respect To} & \textbf{Accessible?} & \textbf{Reflective?}
        \\ 
        \midrule
        Segal & $\brc{\Horn\hookrightarrow\ICh{\Two}}$ & Yes & Yes
        \\ 
        Based Segal & $\brc{\Horn\hookrightarrow\ICh{\Two}}$ in $\Type/\II$ or $\Compr{\IsT{j}_\bot\hookrightarrow \II/j}{j:\II}$ & Yes & Yes
        \\
        Rezk & $\brc{\WalkingBiinv\to\One}$ & Yes & Yes
        \\ 
        Sierpi\'nski & $\Compr{\sigma_X\colon X_\bot\to L\prn{X}}{X:\Type}$
        & No(?)
        & (??)
        \\ 
        Replete & $\Compr{f\colon A\to B}{\text{$\II$ is $\brc{f}$-local}}$
        & No(?) & Yes
        \\ 
        \bottomrule
      \end{tabular}
    \end{center}
    \caption{Summary of important local classes of types.}
  \end{table*}

  \begin{xsect}[sec:segal-completness]{Segal completeness for composing paths}
    A type $X$ shall be called \emph{Segal complete} when it is local with respect to the inclusion $\Horn\hookrightarrow\ICh{\Two}$: this means that every inner horn $x\rightsquigarrow y \rightsquigarrow z$ in $X$ extends uniquely to a triangle with base $x\rightsquigarrow z$ representing the composite of the two paths:
    \[ 
      \begin{tikzcd}
        \Horn \arrow[r,"\forall"] \arrow[d,hookrightarrow]
        &
        X
        \\ 
        \ICh{\Two}
        \arrow[ur,dashed,swap,sloped,"\exists!"]
      \end{tikzcd}
    \] 

    Riehl and Shulman~\cite{riehl-shulman:2017} show that all higher associativities, \etc, can be obtained from the Segal completeness law. Because Segal completeness can be defined by localising at a small collection of maps, we have a localisation whose reflection is easily computed as a higher inductive type as explained by Rijke~\etal\cite{rijke-shulman-spitters:2020}.

    \begin{remark}
      This presentation of Segal completeness in terms of internal orthogonality is actually older than \opcit, originating in 1990s synthetic domain theory under the name \emph{path transitivity}~\cite{fiore-rosolini:1997:cpos}.
    \end{remark}

    Segal completeness is a higher-dimensional analogue to the concept of a precategory (non-univalent category) in univalent foundations~\cite{hottbook}; in higher category theory, this is referred to as a \emph{flagged} $\infty$-category. What is missing is any relationship between the $\infty$-groupoid core of the $\infty$-category and the underlying $\infty$-groupoid of objects, which are \emph{a priori} different; this gap is addressed by the \emph{Rezk completeness} law in \S~\ref{sec:rezk-completeness}.
  \end{xsect}

  \begin{xsect}[sec:based-segal-completeness]{Based Segal completeness}
    Whereas the Segal completeness law asks for a type $X$ to be local with respect to $\Horn\hookrightarrow\ICh{\Two}$, we can consider an even stronger condition that requires $X$ to be local with respect to the components of $\Horn\hookrightarrow\ICh{\Two}$ over $\II$ as displayed in \S~\ref{sec:display-of-inner-horn-inclusion}. Given $j:\II$ we can explicitly describe these components and the corresponding unique extension law as follows:
    \[ 
      \begin{tikzcd}
        X
        &
        \IsT{j}_\bot 
          \arrow[r,hookrightarrow]
          \arrow[d,hookrightarrow]
          \arrow[l,swap,"\forall"]
          \arrow[dr,phantom,very near start,"\lrcorner"]
        & 
        \Horn
          \arrow[d,hookrightarrow]
        \\ 
        &
        \II/j 
          \arrow[r,hookrightarrow]
          \arrow[d]
          \arrow[dr,phantom,very near start,"\lrcorner"]
          \arrow[ul,dashed,swap,sloped,"\exists!"]
        & 
        \ICh{\Two}
          \arrow[d,"\pi_2"]
        \\ 
        &
        \brc{j} 
          \arrow[r,hookrightarrow]
        &
        \II
      \end{tikzcd}
    \] 

    In other words, a type $X$ is \emph{based Segal complete} if and only if either of the following equivalent conditions hold: 
    \begin{enumerate}
      \item $\II\times X$ is $\brc{\Horn\hookrightarrow\ICh{\Two}}$-local in the slice $\Type/\II$.
      \item $X$ is $\Compr{\IsT{j}_\bot\hookrightarrow \II/j}{j:\II}$-local.
    \end{enumerate}

    \begin{corollary}
      Any based Segal complete type is Segal complete.
    \end{corollary}

  \end{xsect}

  \begin{xsect}[sec:rezk-completeness]{Rezk completeness for contracting equivalences}
    A type $X$ is called \emph{Rezk complete} by Riehl and Shulman~\cite{riehl-shulman:2017} when the inclusion of identity paths in $X$ into equivalences is an isomorphism. In terms of the geometry that we have developed in \S~\ref{sec:equivalences}, this is to say that $X$ is Rezk complete when it is local with respect to the terminal map $\WalkingBiinv\to\One$:
    \[ 
      \begin{tikzcd}
        \WalkingBiinv \arrow[r,"\forall"] \arrow[d]
        &
        X
        \\ 
        \One
        \arrow[ur,dashed,swap,sloped,"\exists!"]
      \end{tikzcd}
    \] 

    In the terminology of Rijke, Shulman, and Spitters~\cite{rijke-shulman-spitters:2020}, Rezk completeness is the \emph{nullification at the walking equivalence}. As such, Rezk completeness forms not only a reflective subuniverse but a \emph{higher modality} in the sense that Rezk complete types are closed under internal sums.
  \end{xsect}

  \begin{xsect}{Sierpi\'nski completeness for analysis of partial elements}
    In the traditional domain theory of dcpos, it is possible to define a map out of $L\prn{Z}$ by ``cases'' on whether the given partial element is defined or not, so long as there is an inequality from the undefined case to the defined case. Although this may appear at first to be an instance of classical reasoning, it is really no more than the dual universal property of the partial map classifier as the Sierpi\'nski cone---which does hold for dcpos even in the constructive setting~\cite{sterling:2024:lifting}.

    In our synthetic setting, so long as $\II$ is consistent, we do have a comparison map $\sigma_Z\colon Z_\bot\to L\prn{Z}$ but we have seen that this map is not likely to be an isomorphism. We can, however, consider a \emph{local class} within which these maps are forced to be isomorphisms. In particular, we define a type $X$ to be \emph{Sierpi\'nski complete} when it is local with respect to every $\sigma_Z\colon Z_\bot\to L\prn{Z}$:
    \[ 
      \begin{tikzcd}
        Z_\bot \arrow[r,"\forall"] \arrow[d,swap,"\sigma_Z"]
        &
        X
        \\ 
        L\prn{Z}
        \arrow[ur,dashed,swap,sloped,"\exists!"]
      \end{tikzcd}
    \] 

    Sierpi\'nski completeness is a very large property, and so there is little hope of the local class being accessible in the sense of Rijke~\etal~\cite{rijke-shulman-spitters:2020}---and existing tools like those of Christensen~\cite{christensen:2024} for obtaining non-accessible localisations do not seem to apply. Later on we will see that there are multiple more restrictive localisations yielding only Sierpi\'nski complete types (but, probably, not all of the Sierpi\'nski complete types).

    The significance of finding a localisation in which all types are Sierpi\'nski complete is that within this subuniverse, the Sierpi\'nski cone really can be computed as a partial map classifier---with the caveat that the Sierpi\'nski cone is therefore not preserved by the inclusion into the ambient universe.
  \end{xsect}

  \begin{xsect}[sec:repleteness]{Repleteness for extending properties of the interval}
    Finally, we consider the most extreme localisation possible that includes the interval: Hyland's \emph{replete types}~\cite{hyland:1991}. A type $X$ is said to be replete when it is local for the class of maps $f\colon A \to B$ such that $\II$ is $\brc{f}$-local. In other words, $X$ ``sees'' only that which the interval sees and thus lies in \emph{every} localisation containing $\II$.
    Although repleteness is described by an extremely large localising class, it is reflective as pointed out Hyland~\cite[Theorem~6.1.1]{hyland:1991}. An inductive construction of the repletion is also given by Streicher~\cite{streicher:1999}.

    Synthetic domain theorists in the tradition of Hyland, Phoa, and Taylor have relied on repleteness as a simple way to get a category of predomains satisfying as many desirable conditions as possible whilst remaining non-trivial. Of course, because $\II$ is a set in the sense of univalent founations, so is any replete type; this is why a higher-dimensional domain theorist might wish to look beyond the replete objects, although admittedly our main result (Theorem~\ref{thm:main-result}) applies only to sets.
  \end{xsect}

\begin{xsect}[sec:fiore-lemma]{Generalised well-completeness and Fiore's lemma}
  We recall that $\OpenEmbeddings$ is the class of all open embeddings; given a class of maps $\mathcal{L}$, we shall write $\OpenEmbeddings^*\mathcal{L}$ to be the closure of $\mathcal{L}$ under pullback along open embeddings.
  The following is an old result that we learned from Marcelo Fiore.

  \begin{lemma}[Fiore]\label{lem:fiore-lemma}
    Let $\mathcal{L}$ be a class of maps for which $\II$ is $\mathcal{L}$-local; if $A$ is $\OpenEmbeddings^*\mathcal{L}$-local, then $L\prn{A}$ is $\mathcal{L}$-local.
  \end{lemma}

  \begin{convention}[Well-completeness]
    In synthetic domain theory, a type is called \emph{complete} when it is local with respect to the comparison map from the initial $L$-algebra to the final $L$-coalgebra;
    Longley~\cite{longley:1995} called a type $X$ \emph{well-complete} when $L\prn{X}$ is complete. By the same token, we shall use the following terminologies:
    \begin{quote}
      A type $X$ is (Segal, based Segal, Rezk, Sierpi\'nski) \emph{well-complete} when $L\prn{X}$ is (Segal, based Segal, Rezk, Sierpi\'nski) complete.
    \end{quote}
  \end{convention}

  Lemma~\ref{lem:fiore-lemma} gives us the blueprint for developing local classes that are closed under partial map classifiers: either we prove that a given local class is stable under pullback along open embeddings, or we replace it with one that is.
\end{xsect}
 \end{xsect}

\begin{xsect}[sec:phoa]{Phoa's principle}
  A standard axiom of synthetic domain theory is the \emph{Phoa principle}, which can be stated with respect to an arbitrary interval $\II$.

  \begin{definition}[Phoa principle]\label{def:phoa}
    An interval $\II$ satisfies the \emph{Phoa principle} when either of the following equivalent conditions hold:
    \begin{enumerate}
      \item Every function $\alpha\colon \II\to\II$ is monotone and the embedding $\brc{0,1}\hookrightarrow\II$ is an $\II$-epimorphism.
      \item The boundary map $\II^{\prn{1,0}}\colon \II^\II \to \II^2$ is an embedding with image $\ICh{\Two}\hookrightarrow \II^2$, so we have $\II^\II\cong \ICh{\Two}$. 
    \end{enumerate}
  \end{definition}

  \begin{xsect}{Phoa's principle for chains}
    The Phoa principle can be generalised to chains.

    \begin{definition}
      An interval $\II$ satisfies the \emph{$n$-dimensional Phoa principle} when either of the following equivalent conditions hold:
      \begin{enumerate}
        \item Every function $\alpha\colon \ICh{n}\to\II$ is monotone and the embedding $\Compr{\prn{1^k\ldots 0^{n-k}}}{0\leq k \leq n}\hookrightarrow \ICh{n}$ is an $\II$-epimorphism.
        \item The boundary map $\II^{\ICh{n}}\to\II^{n+1}$ is an embedding with image $\ICh{n+1}\hookrightarrow \II^{n+1}$, so we have $\II^{\ICh{n}}\cong \ICh{n+1}$.
      \end{enumerate}
    \end{definition}

    Thus the ordinary Phoa principle is precisely the $1$-dimensional Phoa principle; of course, the $0$-dimensional Phoa principle is always true.

    \begin{lemma}\label{lem:bdl-phoa}
      When $\II$ is a bounded distributive lattice, the following are equivalent:
      \begin{enumerate}
        \item The Phoa principle holds.
        \item The $n$-dimensional Phoa principle holds for all $n$.
        \item For any $\alpha\colon \II\to\II$ we have the ``linear interpolation'' equation $\alpha\prn{i}=\alpha\prn{0}\sqcup\prn{i\sqcap\alpha\prn{1}}$.
        \item For any $\alpha\colon\ICh{n}\to\II$ we have $\alpha\prn{\vec\imath} = \alpha\prn{0^n\ldots} \sqcup \bigsqcup_{\prn{1\leq k\leq n}}\prn{\vec\imath_{n-k} \sqcap \alpha\prn{1^k\ldots 0^{n-k}} }$
      \end{enumerate}
    \end{lemma}
  \end{xsect}

  \begin{xsect}{Phoa's principle on slices of the interval}
    For any $i:\II$, we shall write $\II/i$ for the slice of $\II$ viewed as a partial order; geometrically, we think of $\II/i$ as the interval with $i$ and everything above it smooshed onto the top endpoint. Abusing notation slightly, we shall $\ICh{n}/i$ for the subposet of $\ICh{n}$ spanned by chain whose vertices are all in $\II/i$. Then the ordinary Phoa principle then implies a relative version for the smooshed interval:

    \begin{lemma}[Relative Phoa principle]\label{lem:relative-phoa-principle}
      Let $\II$ be an interval satisfying the Phoa principle. Then for any $i:\II$, every function $\alpha\colon \II/i\to \II$ is monotone and the embedding $\brc{0,i}\hookrightarrow \II$ is an $\II/i$-epimorphism.
    \end{lemma}
    \begin{IEEEproof}
      Fix $i:\II$.  For monotonicity, we fix a function $\alpha\colon \II/i\to \II$. We define $\hat{\alpha}\colon \II\to \II$ to be restriction of $\alpha$ along the meet functor $i^*\colon \II\to \II/i$; explicitly, we have $\hat{\alpha}\prn{j} :\equiv \alpha\prn{j\sqcap i}$. To show that $\alpha$ is monotone, we fix $j\sqsubseteq k\sqsubseteq i$ to show that $\alpha\prn{j} \sqsubseteq \alpha\prn{k}$. By the ordinary Phoa principle, we have $\alpha\prn{j}=\alpha\prn{j\sqcap i} \equiv \hat{\alpha}\prn{j} \sqsubseteq \hat{\alpha}\prn{k} \equiv \alpha\prn{k\sqcap i} = \alpha\prn{i}$.

      To show that $\brc{0,i}\hookrightarrow \II$ is an $\II/i$-epimorphism, we must show that for any $\alpha,\beta\colon \II/i \to \II$, if $\alpha\prn{0}=\beta\prn{0}$ and $\alpha\prn{i}=\beta\prn{i}$ then we have $\alpha=\beta$. It suffices to show that $\hat{\alpha}=\hat{\beta}$, as for any $j:\II/i$ we have $\hat\alpha\prn{j}=\alpha\prn{j}$. By the ordinary Phoa principle, we have $\hat\alpha=\hat{\beta}$ because $\hat\alpha\prn{0} = \alpha\prn{0} = \beta\prn{0} = \hat\beta\prn{0}$ and $\hat\alpha\prn{1}=\alpha\prn{i} = \beta\prn{i}=\hat{\beta}\prn{1}$.
    \end{IEEEproof}

    \begin{corollary}
      When $\II$ is a bounded distributive lattice satisfying the Phoa principle, for any $\alpha\colon \II/i\to \II$ and $j\sqsubseteq i$ we have $\alpha\prn{j} = \alpha\prn{0}\sqcup \prn{j\sqcap \alpha\prn{i}}$.
    \end{corollary}
  \end{xsect}

  \begin{xsect}[sec:phoa-consequences]{Consequences of Phoa's principle}
    Phoa's principle is the main ingredient to ensure that the interval lies in various important local classes; in particular, we will show in this section that under appropriate assumptions, the interval is based Segal complete, Segal complete, and Rezk complete. We do not derive Segal completeness from based Segal completeness, because the two lemmas seem to require incomparable assumptions.

    \begin{theorem}\label{thm:interval-is-based-segal-complete}
      Suppose that $\II$ is a consistent, bounded distributive lattice satisfying the Phoa principle, and is equipped with an internal sum structure that factors binary meets in the sense of Definition~\ref{def:factoring-meets}. Then $\II$ is based Segal complete.
    \end{theorem}

    \begin{IEEEproof}
      Fixing $j:\II$ we must check that the interval is $\brc{\IsT{j}_\bot\hookrightarrow \II/j}$-local in the following sense:
      \[ 
        \begin{tikzcd}
          \IsT{j}_\bot
            \arrow[r,"\forall\alpha"]
            \arrow[d,hookrightarrow]
          &
          \II
          \\ 
          \II/j
            \arrow[ur,dashed,swap,sloped,"\exists!\hat\alpha"]
        \end{tikzcd}
      \]

      We define the extension $\hat\alpha\colon\II/j\to\II$ as follows, recalling that $\IsT{j}_\bot = \Compr{i:\II}{\IsF{i}\lor\IsT{j}}$:
      \[
        \textstyle
        \hat\alpha\prn{j\sqsupseteq i} :\equiv 
        \alpha\prn{0}
        \sqcup 
        \sum\prn{i,\Lam{p:\IsT{i}}\alpha\prn{j}}
      \] 

      It was necessary above to use the internal sum structure because the application $\alpha\prn{j}$ makes sense only when $\IsF{i}\lor\IsT{j}$, which holds under $\IsT{i}$ because $i\sqsubseteq j$.

      We fix $i:\II$ such that $\IsF{i}\lor\IsT{j}$ holds to check $\hat\alpha\prn{i} = \alpha\prn{i}$. We first consider the case where $\IsF{i}$ holds:
      \begin{align*}
        \IsF{i} &\vdash 
        \hat\alpha\prn{i}
        \\ 
        &= 
        \textstyle
        \alpha\prn{0}
        \sqcup 
        \sum\prn{0,\Lam{p:\IsT{0}}\alpha\prn{j}}
        &&\text{by definition}
        \\ 
        &=
        \textstyle
        \alpha\prn{0}
        \sqcup 
        \sum\prn{0,\Lam{\_:\IsT{0}}0} 
        &&\text{$\II$ is consistent}
        \\ 
        &= 
        \alpha\prn{0}\sqcup \prn{0\sqcap 0}
        &&\text{$\textstyle\sum$ factors $\sqcap$}
        \\ 
        &= \alpha\prn{i}
      \end{align*}

      On the other hand, if $\IsT{j}$ holds then both $\IsT{j}_\bot \cong \II$ and $\II/j\cong\II$ and thus under these identifications, the ordinary Phoa principle applies to maps $\IsT{j}_\bot\to\II$. Thus it remains only to check that $\hat\alpha\prn{1} = \alpha\prn{1}$.
      \begin{align*}
        \IsT{j} &\vdash 
        \hat\alpha\prn{1}
        \\ 
        &= 
        \textstyle
        \alpha\prn{0} \sqcup \sum\prn{1,\Lam{\_:\IsT{1}} \alpha\prn{1}}
        &&\text{by definition}
        \\ 
        &= 
        \alpha\prn{0}\sqcup \prn{1\sqcap \alpha\prn{1}}
        &&\text{$\textstyle\sum$ factors $\sqcap$}
        \\ 
        &= \alpha\prn{0}\sqcup\alpha\prn{1}
        &&\text{by algebra}
        \\ 
        &= \alpha\prn{1}
        &&\text{by Phoa's principle}
      \end{align*}

      In the last step, we used the fact that $\alpha$ is monotone by Phoa's principle, and so $\alpha\prn{0}\sqsubseteq\alpha\prn{1}$ and thus $\alpha\prn{0}\sqcup \alpha\prn{1} = \alpha\prn{1}$.

      We must show that any further extension $\beta\colon \II/j\to \II$ of $\alpha\colon\IsT{j}_\bot\to\II$ is equal to $\hat\alpha$. Fixing $i\sqsubseteq j$, we proceed as follows:
      \begin{align*}
        \beta\prn{i} 
        &= \beta\prn{0}\sqcup\prn{i\sqcap \beta\prn{j}}
        &&\text{by Phoa's principle}
        \\ 
        &= 
        \textstyle
        \beta\prn{0}\sqcup \sum\prn{i,\Lam{\_:\IsT{i}} \beta\prn{j}}
        &&\text{$\textstyle\sum$ extends $\sqcap$}
        \\ 
        &=
        \textstyle
        \alpha\prn{0}\sqcup \sum\prn{i,\Lam{p:\IsT{i}} \alpha\prn{j}}
        &&\text{$\beta$ extends $\alpha$ and $i\sqsubseteq j$}
        \\
        &= 
        \hat{\alpha}\prn{i}
        &&\text{by definition}
      \end{align*}
      
      Thus we conclude that $\hat\alpha\colon \II/j\to \II$ is the unique extension of $\alpha\colon\IsT{j}_\bot\to\II$ along $\IsT{j}_\bot\hookrightarrow\II/j$.
    \end{IEEEproof}

    \begin{corollary}
      If $\II$ is a dominance closed under finite disjunctions and satisfying the Phoa principle, then $\II$ is based Segal complete.
    \end{corollary}

    \begin{theorem}\label{thm:interval-is-segal-complete}
      Any interval $\II$ that satisfies the Phoa principle in dimensions $n \leq 2$ is Segal complete.
    \end{theorem}
  
    \begin{IEEEproof}
      We must show that the comparison map $\II^\eta\colon \II^{\ICh{\Two}}\to \II^{\Horn}$ induced by the inclusion $\eta\colon \Horn \hookrightarrow \ICh{\Two}$ is an isomorphism.
      By the universal property of the pushout square that defines the horn $\Horn$, the following is a pullback square:
      \[
        \begin{tikzcd}
          \II^{\Horn}
            \arrow[d,swap, "\II\Sup{\prn{-\sqsupseteq0}}"]
            \arrow[r,"\II\Sup{\prn{1\sqsupseteq -}}"]
            \arrow[dr,phantom,very near start, "\lrcorner"] 
          & 
          \II^\II
            \arrow[d,"\II^0"]
          \\ 
          \II^\II
            \arrow[r,swap,"\II^1"] 
          & 
          \II
        \end{tikzcd}
      \]
      
      By the Phoa principle, then, each of the following squares is cartesian and so the outer square is cartesian.
      \[
        \begin{tikzcd}[column sep = large]
          \II^{\Horn}
            \arrow[d,swap, "\II\Sup{\prn{-\sqsupseteq0}}"]
            \arrow[r,"\II\Sup{\prn{1\sqsupseteq-}}"]
            \arrow[dr,phantom,very near start,"\lrcorner"] 
          & 
          \II^\II
            \arrow[d,"\II^0" description]
            \arrow[r,"\prn{\II^1\sqsupseteq \II^0}"]
            \arrow[dr,phantom,very near start,"\lrcorner"] 
          & 
          \ICh{\Two}
            \arrow[d,"\pi_1"]
          \\ 
          \II^\II
            \arrow[r,"\II^1" description] 
            \arrow[d,swap,"\prn{\II^1\sqsupseteq \II^0}"]
            \arrow[dr,phantom,very near start,"\lrcorner"]
          & 
          \II
            \arrow[r,equals]
            \arrow[d,equals]
            \arrow[dr,phantom,very near start,"\lrcorner"] 
          & 
          \II
            \arrow[d,equals]
          \\ 
          \ICh{\Two} 
            \arrow[r,swap,"\pi_2"]
          &
          \II 
            \arrow[r,equals]
          &
          \II
        \end{tikzcd}
      \]
      
      In other words, an element of $\II^{\Horn}$ is given precisely by a chain $\prn{k\sqsupseteq j \sqsupseteq i}$. As $\ICh{\Three}$ is by definition \emph{also} the pullback of the same co-span, we can define a unique isomorphism $\psi\colon \II^{\Horn}\to \ICh{\Three}$ factoring like so:
      \[ 
        \begin{tikzcd}[column sep=large]
          \II^\II
            \arrow[d,swap,"\prn{\II^1\sqsupseteq\II^0}"]
          &
          \II^{\Horn}
            \arrow[r,"\II\Sup{\prn{1\sqsupseteq-}}"]
            \arrow[d,"\psi"description]
            \arrow[l,swap,"\II\Sup{\prn{-\sqsupseteq0}}"]
          & 
          \II^\II
            \arrow[d,"\prn{\II^1\sqsupseteq\II^0}"]
          \\ 
          \ICh{\Two}
          &
          \ICh{\Three}
            \arrow[l,"\prn{\pi_2\sqsupseteq\pi_3}"]
            \arrow[r,swap,"\prn{\pi_1\sqsupseteq\pi_2}"]
          &
          \ICh{\Two}
        \end{tikzcd}
      \]
      
      If we define $\psi\prn{\alpha} :\equiv \prn{\alpha\prn{1\sqsupseteq 1}\sqsupseteq\alpha\prn{1\sqsupseteq 0}\sqsupseteq\alpha\prn{0\sqsupseteq 0}}$, the squares above commute definitionally.
      With this definition, we see that the isomorphism $\psi\colon \II^{\Horn}\to\ICh{\Three}$ factors the Phoa isomorphism $\phi\colon \II^{\ICh{\Two}}\to\ICh{\Three}$ \emph{definitionally} through the comparison map $\II^\eta\colon \II^{\ICh{\Two}}\to\II^{\Horn}$ as depicted below: 
      \[ 
        \begin{tikzcd}
          \II^{\ICh{\Two}} 
            \arrow[rr,"\phi"]
            \arrow[dr,sloped,swap,"\II^\eta"]
          && 
          \ICh{\Three}
          \\ 
          &
          \II^{\Horn}
            \arrow[ur,swap,sloped,"\psi"]
        \end{tikzcd}
      \]
      
      Therefore, by the three-for-two principle of isomorphisms, the comparison map $\II^\eta\colon \II^{\ICh{\Two}}\to\II^{\Horn}$ is an isomorphism.
    \end{IEEEproof}
  
    \begin{lemma}\label{lem:phoa-principle-makes-interval-path-univalent}
      Any interval $\II$ satisfying the Phoa principle in dimensions $n\leq 2$ is Rezk complete.
    \end{lemma}
  
    \begin{IEEEproof}
      Because $\II$ satisfies the Phoa principle by assumption and is thus Segal complete by Theorem~\ref{thm:interval-is-segal-complete}, an element of $\II^\WalkingBiinv$ consists precisely of two points $i,j:\II$ together a witness $f\colon j\sqsupseteq i$ and a section and retraction $s_f,r_f\colon i\sqsupseteq j$. This proves that $i=j$ because $\sqsupseteq$ is a partial order; the remaining data is propositional because $\II$ is a set.
    \end{IEEEproof}

  \end{xsect}

\begin{xsect}{Closure of local classes under partial map classifiers}
  In \S~\ref{sec:phoa-consequences} we showed that the Phoa principle implies that several important local classes (Rezk, Segal, and based Segal) contain the interval under appropriate assumptions. In this section, we shall investigate closure of these local classes under partial map classifiers.

  \begin{lemma}\label{lem:rezk-left-class-stable}
    Let $\II$ satisfy the Phoa principle in dimensions $n\leq 2$. A type $X$ is Rezk complete if and only if it is Rezk well-complete.
  \end{lemma}

  \begin{IEEEproof}
    Only the forward direction is non-trivial.
    We apply Lemma~\ref{lem:fiore-lemma}, which requires that $\II$ be Rezk complete (Lemma~\ref{lem:phoa-principle-makes-interval-path-univalent}) and that $X$ be local with respect to the 
    the pullback of $\WalkingBiinv\to\One$ along an open embedding $\IsT{i}\hookrightarrow\One$. This pullback is the product $\IsT{i}\times \WalkingBiinv\to \IsT{i}\times\One$, and the left class of \emph{any} internal localisation is closed under such products.
  \end{IEEEproof}

  \begin{lemma}\label{thm:segal-complete-vs-well-complete}
    Suppose that $\II$ is a bounded distributive lattice satisfying the disjunction property and the Phoa principle. Then a type $X$ is Segal complete if and only if it is Segal well-complete.
  \end{lemma}

  \begin{IEEEproof}
    Only the forward direction is difficult; in this case, we apply Lemma~\ref{lem:fiore-lemma} which requires that $\II$ be Segal complete and that $X$ be local with respect to the restriction of the horn inclusion along an open embedding. For the former, recall from Lemma~\ref{lem:bdl-phoa} that the Phoa principle extends to all dimensions because $\II$ is assumed to be a bounded distributive lattice, so we use Theorem~\ref{thm:interval-is-segal-complete} to deduce that $\II$ is Segal complete. For the latter, let $\alpha\colon \ICh{\Two}\to\II$ be any function, and consider the induced embedding $\eta.\alpha\colon\Horn.\alpha\hookrightarrow\ICh{\Two}.\alpha$ as defined below:
    \[ 
      \begin{tikzcd}
        \Horn.\alpha
          \arrow[d,hookrightarrow]
          \arrow[r,hookrightarrow,"\eta.\alpha"]
          \arrow[dr,phantom,very near start,"\lrcorner"]
        &
        \ICh{\Two}.\alpha
          \arrow[r]
          \arrow[d,hookrightarrow]
          \arrow[dr,phantom,very near start,"\lrcorner"]
        &
        \brc{1}
          \arrow[d,hookrightarrow]
        \\ 
        \Horn 
          \arrow[r,hookrightarrow,swap,"\eta"]
        &
        \ICh{\Two}
          \arrow[r,swap,"\alpha"] 
        &
        \II
      \end{tikzcd}
    \]

    We wish to show that $X$ is $\eta.\alpha$-local. We shall abbreviate $\alpha\prn{j\sqsupseteq i}$ by $\alpha_{ji}$.
    By the Phoa principle, we have $\alpha_{ji} = \alpha_{00}\sqcup \beta_{ji}$ where $\beta_{ji} :\equiv \prn{j\sqcap \alpha_{10}}\sqcup \prn{i \sqcap \alpha_{11}}$. Because $\II$ is assumed to satisfy the disjunction property and so $\IsT{-}\colon\II\to\Prop$ preserves binary joins, we can give an explicit description of the embedding $\eta.\alpha\colon\Horn.\alpha\hookrightarrow\ICh{\Two}.\alpha$.
    \begin{align*}
      \Horn.\alpha &= 
      \Compr{\prn{j\sqsupseteq i}\in \Horn}{
        \IsT{\alpha_{00}}\lor 
        \IsT{\beta_{ji}}
      }
      \\
      \ICh{\Two}.\alpha &= 
      \Compr{\prn{j\sqsupseteq i}\in \ICh{\Two}}{
        \IsT{\alpha_{00}}\lor \IsT{\beta_{ji}}
      }
    \end{align*}
    
    We start by visualising the descriptions above as pushout squares, which we can compute as follows. First we compute the \emph{conjunction} $\IsT{\alpha_{00}}\land\IsT{\beta_{ji}}$ for any $\prn{j\sqsupseteq i}$:
    \begin{align*}
      &\IsT{\alpha_{00}}\land\IsT{\beta_{ji}} 
      \\ 
      &\quad= 
      \IsT{\alpha_{00}}\land\prn{
        \prn{\IsT{j}\land\IsT{\alpha_{10}}}
        \lor 
        \prn{\IsT{i}\land\IsT{\alpha_{11}}}
      }
      \\ 
      &\quad\Leftrightarrow
      \prn{\IsT{j}\land\IsT{\alpha_{00}}\land\IsT{\alpha_{10}}}
      \lor \prn{\IsT{i}\land\IsT{\alpha_{00}}\land\IsT{\alpha_{11}}}
      \\ 
      &\quad\Leftrightarrow
      \prn{\IsT{j}\land\IsT{\alpha_{00}}}
      \lor \prn{\IsT{i}\land\IsT{\alpha_{00}}}
      \\
      &\quad\Leftrightarrow
      \IsT{\alpha_{00}}\land\prn{\IsT{i}\lor\IsT{j}}
      \\
      &\quad\Leftrightarrow
      \IsT{\alpha_{00}}\land\IsT{j}
    \end{align*}

    We note that for any $\prn{j\sqsupseteq i}\in \ICh{\Two}$, if $\IsT{i}\lor\IsT{j}$ holds then we certainly have $\prn{j\sqsupseteq i}\in\Horn$. Therefore, the canonical embedding $\eta.\beta\colon \Horn.\beta\hookrightarrow \ICh{\Two}.\beta$ is an isomorphism. We therefore have the following pushout squares:
    \[ 
      \begin{tikzcd}[cramped]
        \ICh{\Two}\times\IsT{\alpha_{00}}
          \arrow[d,hookrightarrow]
        &
        \brc{\prn{1\sqsupseteq i}:\Horn}\times
        \IsT{\alpha_{00}}
          \arrow[d,hookrightarrow]
          \arrow[r,hookrightarrow]
          \arrow[l,hookrightarrow]
          \arrow[dr,phantom,very near end, "\ulcorner"]
          \arrow[dl,phantom,very near end, "\urcorner"]
        & 
        \Horn\times\IsT{\alpha_{00}}
          \arrow[d,hookrightarrow]
        \\
        \ICh{\Two}.\alpha
        &
        \Horn.\beta
          \arrow[r,hookrightarrow]
          \arrow[l,hookrightarrow]
        &
        \Horn.\alpha
      \end{tikzcd}
    \]

    By the universal property of the left-hand square, the following is a pullback square, as $\brc{\prn{1\sqsupseteq i}:\Horn}$ is just $\II$:
    \[ 
      \begin{tikzcd}
        X^{\ICh{\Two}.\alpha}
          \arrow[r]
          \arrow[d]
          \arrow[dr,phantom,very near start,"\lrcorner"]
        &
        X\Sup{\ICh{\Two}\times\IsT{\alpha_{00}}}
          \arrow[d,"X\Sup{\prn{1\sqsupseteq-}\times\IsT{\alpha_{00}}}"]
        \\
        X\Sup{\ICh{\Two}.\beta}
          \arrow[r,swap,"X\Sup{\prn{1\sqsupseteq-}}"]
        &
        X\Sup{\II\times\IsT{\alpha_{00}}}
      \end{tikzcd}
    \]
    
    By assumption on $X$ we know that the restriction map $X\Sup{\eta\times\IsT{\alpha_{00}}}\colon X\Sup{\ICh{\Two}\times\IsT{\alpha_{00}}}\to X\Sup{\Horn\times\IsT{\alpha_{00}}}$ is an isomorphism. Therefore, the following composite square is cartesian:
    \[ 
      \begin{tikzcd}[row sep=large, column sep=large]
        X^{\ICh{\Two}.\alpha}
          \arrow[r]
          \arrow[d]
          \arrow[dr,phantom,very near start,"\lrcorner"]
        &
        X\Sup{\ICh{\Two}\times\IsT{\alpha_{00}}}
          \arrow[d,"X\Sup{\prn{1\sqsupseteq-}\times\IsT{\alpha_{00}}}" description]
          \arrow[r,"X\Sup{\eta\times\IsT{\alpha_{00}}}"]
          \arrow[dr,phantom,very near start,"\lrcorner"]
        &
        X\Sup{\Horn\times\IsT{\alpha_{00}}}
          \arrow[d,"X\Sup{\prn{1\sqsupseteq-}\times\IsT{\alpha_{00}}}"]
        \\
        X\Sup{\ICh{\Two}.\beta}
          \arrow[r,"X\Sup{\prn{1\sqsupseteq-}}"description]
          \arrow[d,swap,"X\Sup{\eta.\beta}"]
          \arrow[dr,phantom,very near start,"\lrcorner"]
        &
        X\Sup{\II\times\IsT{\alpha_{00}}}
          \arrow[r,equals]
          \arrow[d,equals]
          \arrow[dr,phantom,very near start,"\lrcorner"]
        &
        X\Sup{\II\times\IsT{\alpha_{00}}}
          \arrow[d,equals]
        \\
        X\Sup{\Horn.\beta}
          \arrow[r,swap,"X\Sup{\prn{1\sqsupseteq-}}"]
        &
        X\Sup{\II\times\IsT{\alpha_{00}}}
          \arrow[r,equals]
        &
        X\Sup{\II\times\IsT{\alpha_{00}}}
      \end{tikzcd}
    \] 

    By definition, $X\Sup{\Horn.\alpha}$ is the pullback of the same co-span and, moreover, it can be seen that the gap isomorphism $X\Sup{\ICh{\Two}.\alpha}\to X\Sup{\Horn.\alpha}$ induced by the two pullback squares is precisely the restriction map $X^{\eta.\alpha}$.
  \end{IEEEproof}

  \begin{theorem}\label{thm:based-segal-complete-vs-well-complete}
    Suppose that $\II$ is a consistent bounded distributive lattice satisfying the disjunction property and the Phoa principle, equipped with an internal sum structure that factors binary meets. Then a type $X$ is \emph{based} Segal complete if and only if it is \emph{based} Segal well-complete.
  \end{theorem}

  The proof is analogous to that of Theorem~\ref{thm:segal-complete-vs-well-complete}.

  Riehl and Shulman~\cite{riehl-shulman:2017} defined a (synthetic) \emph{$\infty$-category} to be a type that is both Segal and Rezk complete. Our results so far show that under reasonable assumptions, the synthetic $\infty$-categories (as well as the stronger version with the \emph{based} Segal condition) are closed under partial map classifiers.

  \begin{corollary}\label{cor:infty-cat-pmc}
    When $\II$ is a bounded distributive lattice satisfying the disjunction property (\eg a total order) and the Phoa principle, if $X$ is an $\infty$-category then so is $L\prn{X}$.
  \end{corollary}

  \begin{remark}
    The assumptions of Theorems~\ref{thm:segal-complete-vs-well-complete}~and~\ref{thm:based-segal-complete-vs-well-complete} (that $\II$ is a bounded distributive lattice satisfying the disjunction principle and the Phoa principle) are not too strong, but even if these assumptions do not hold, one can always obtain a localisation closed under partial map classifiers by stabilising the left class under pullback along open embeddings. That is the force of Fiore's lemma (Lemma~\ref{lem:fiore-lemma}).
  \end{remark}
\end{xsect}
 
\end{xsect}
\begin{xsect}{Main results on Sierpi\'nski completeness}
  We noted earlier that Sierpi\'nski complete types do not obviously form an (internal) localisation because the left class is too large. We will now describe a more restricted local class that is generated by a small left class and is thus accessible.

  \begin{xsect}{Based Segal vs.\ Sierpi\'nski completeness}
    We can now expound our main results, showing that any based Segal complete set is also Sierpi\'nski complete. 

    \begin{lemma}\label{lem:sierp-complete-implies-based-segal-complete}
      When $\II$ is conservative, any Sierpi\'nski complete type is also based Segal complete.
    \end{lemma}
    \begin{IEEEproof}
      In this case, the canonical inclusion $\II/j\hookrightarrow L\IsT{j}$ over $\II$ is an isomorphism. Moreover, the composite inclusion $\IsT{j}_\bot\hookrightarrow\II/j\cong L\IsT{j}$ is precisely the comparison map $\sigma_{\IsT{j}}\colon \IsT{j}_\bot\to L\IsT{j}$.
    \end{IEEEproof}

    For the remainder of this section, we assume that $\II$ is consistent so that $\sigma_X\colon X_\bot\to L\prn{X}$ exists for each $X$.

    \begin{definition}
      Let $C$ be a based Segal complete type. We define the \emph{synthetic colimit} of a figure $h\colon \IsT{j}_\bot\to C$ to be the element $\bigvee\Sub{\prn{p:\IsT{j}_\bot}}h\prn{p}:C$ defined by evaluating the unique extension of $h\colon \IsT{j}_\bot\to C$ at the top element $j\in\II/j$:
      \[ 
        \begin{tikzcd}
          \IsT{j}_\bot 
            \arrow[d, hookrightarrow]
            \arrow[r, "h"]
          &
          C
          \\ 
          \II/j
            \arrow[ur,sloped,"\hat{h}"description]
          & 
          \brc{j}
            \arrow[l,hookrightarrow]
            \arrow[u,swap,"\bigvee\Sub{\prn{p:\IsT{j}_\bot}}h\prn{p}"]
        \end{tikzcd}
      \] 
    \end{definition}

    \begin{remark}
      The synthetic colimit of a figure $\IsT{j}_\bot\to C$ is precisely the synthetic analogue to the ``$\delta$-joins'' studied by De Jong~\cite{dejong:2023:thesis} in the context of constructive domain theory.
    \end{remark}

    It is not difficult to establish the following lemma using the structure identity principle for $C^{X_\bot}$ developed in \S~\ref{sec:sierpinski-data}.

    \begin{lemma}\label{lem:extension-retraction}
      For any based Segal complete type $C$, 
      each restriction map $C\Sup{\sigma_X}\colon C\Sup{L\prn{X}}\to C\Sup{X_\bot}$ has a retraction sending $f\colon X_\bot\to C$ to the following map $\tilde{f}\colon L\prn{X}\to C$:
      \[ 
        \textstyle
        \tilde{f}\prn{j,x:X\Sup{\IsT{j}}} :\equiv 
        \bigvee\Sub{\prn{p:\IsT{j}_\bot}}f\prn{x_\bot\prn{p}}
      \] 

      We have written $x_\bot\colon \IsT{j}_\bot\to X_\bot$ above for the evident functorial map.
    \end{lemma}

    A converse to Lemma~\ref{lem:extension-retraction} that exhibits a \emph{section} to the restriction maps $C\Sup{\sigma_X}\colon C\Sup{L\prn{X}}\to C\Sup{X_\bot}$ is considerably more difficult, and it is unclear how to achieve it without additional assumptions that force $X$ to be a preorder. The best we could do is Lemma~\ref{lem:extension-section} below:

    \begin{lemma}\label{lem:extension-section}
      When $C$ is a based Segal complete \emph{set}, each $C\Sup{\sigma_X}\colon C\Sup{L\prn{X}}\to C\Sup{X_\bot}$ has a section.
    \end{lemma}

    \begin{IEEEproof}
      We will show that the underlying map of the retraction defined in Lemma~\ref{lem:extension-retraction} is also a section of the restriction map. Recalling the isomorphism $C^{X_\bot}\cong\mathsf{SierpData}_X\prn{C}$ and the structure identity principle for Sierpi\'nski data developed in \S~\ref{sec:sierpinski-data}, we must show the following for each $f\equiv\prn{f^\bot,f^\varsigma,H_f}:\mathsf{SierpData}_X\prn{C}$:
      \begin{enumerate}
        \item We must check $\tilde{f}\prn{0,\mathsf{abort}} = f^\bot$. This follows from the extension property of $C\Sup{\II/0}\to C\Sup{\IsT{0}_\bot}$.
        \item We must check that for each $i:\II$ and $x:X$, we have $\tilde{f}\prn{i,\Lam{\_}x} = f^\varsigma\prn{i,x}$. For this we use both the assumed extension property and its uniqueness law.
        \item We do not need to exhibit the final coherence because $C$ is assumed to be a set.
      \end{enumerate}
      We have finished.
    \end{IEEEproof}

    \begin{theorem}[Main result]\label{thm:main-result}
      Any based Segal complete set is Sierpi\'nski complete.
    \end{theorem}
    \begin{IEEEproof}
      Immediate by Lemmas~\ref{lem:extension-retraction}~and~\ref{lem:extension-section}.
    \end{IEEEproof}

    \begin{remark}
      There does not seem to be any \emph{essential} reason for the assumption that $C$ is a set, but we must admit that we could not manage to prove the necessary coherence without this assumption. A previous version of this paper additionally required boundary separation (which determines paths uniquely by their endpoints), but this assumption turned out to be unnecessary.
    \end{remark}

    \begin{corollary}
      Assuming that the interval is consistent and conservative, a set is Sierpi\'nski complete if and only if it is based Segal complete.
    \end{corollary}

    \begin{IEEEproof}
      By Theorem~\ref{thm:main-result} and Lemma~\ref{lem:sierp-complete-implies-based-segal-complete}; the latter uses conservativity.
    \end{IEEEproof}
  \end{xsect}

  \begin{xsect}{A reflective subuniverse identifying the Sierpi\'nski cone and the partial map classifier}
    In our main result (Theorem~\ref{thm:main-result}), we have shown that any based Segal complete set is Sierpi\'nski complete. We at last investigate the significance of this result by stating the assumptions under which this gives rise to a reflective subuniverse closed under lifting within which the Sierpi\'nski cone and the partial map classifier coincide.

    \begin{corollary}\label{cor:good-reflective-subuniverse}
      Let $\II$ be a consistent bounded distributive lattice satisfying the Phoa principle and the disjunction property, and is moreover equipped with an internal sum structure factoring binary meets. Then (1) the reflective subuniverse $\mathcal{S}\subseteq\Type$ spanned by based Segal complete sets is closed under partial map classifiers, and (2) for each $X\in\mathcal{S}$, the following is a co-comma square in $\mathcal{S}$:
      \[ 
        \begin{tikzcd}[]
          X\arrow[r,equal]\arrow[d,->] & X\arrow[d, hookrightarrow, "\prn{1, \mathsf{const}}"]\\ 
          \brc{\prn{0,\mathsf{abort}}}\arrow[r,hookrightarrow]\arrow[ur,phantom,"\Uparrow"] & L\prn{X}
        \end{tikzcd}
      \] 
      In other words, $L\prn{X}$ is the Sierpi\'nski cone in $\mathcal{S}$.
    \end{corollary}

    (Note that we could have in particular assumed that $\II$ is a \emph{dominance} that satisfies the Phoa principle and is closed under finite disjunctions.)
  
    \begin{IEEEproof}
      The interval is based Segal complete by Theorem~\ref{thm:interval-is-based-segal-complete}, so we have $\II\in\mathcal{S}$. By Theorem~\ref{thm:based-segal-complete-vs-well-complete}, a type $X\in\mathcal{S}$ is based Segal complete if and only if it is based Segal well-complete (here we use the disjunction property), and so we see that $\mathcal{S}$ is closed under partial map classifiers. That the depicted lax square is a co-comma square then follows from Theorem~\ref{thm:main-result}.
    \end{IEEEproof}

    Of course, the reflective subuniverse of replete types from \S~\ref{sec:repleteness} would also satisfy the same properties of Corollary~\ref{cor:good-reflective-subuniverse}; the benefit of our bespoke reflective subuniverse is that it is likely to be strictly larger than the replete types.
  \end{xsect}

\end{xsect} %
\begin{xsect}[sec:conclusion]{Conclusions and future work}
  In Corollary~\ref{cor:good-reflective-subuniverse} we exhibited a reflective subuniverse closed under partial map classifiers, within which the partial map classifier and the Sierpi\'nski cone coincide. Our reflective subuniverse is almost certainly  less restrictive than the replete types, the most extreme reflective subuniverse containing the interval. This is a good first step and is immediately applicable to denotational semantics in synthetic domain theory, but we would still hope to find a broader localisation containing untruncated spaces. Future work that analyses the Sierpi\'nski cone construction for synthetic higher categories may therefore require some new ideas.

  We have left unexplored an important connection between our assumptions and the \emph{duality} or \emph{synthetic quasicoherence} axiom of Blechschmidt~\cite{blechschmidt:2017}, which Gratzer~\etal\cite{gratzer-weinberger-buchholtz:2024} have recently employed in the context of synthetic higher categories. In our context, synthetic quasicoherence for the generic distributive lattice implies conservativity, Phoa's principle, and the dominance property. There is much more to say here that we must defer to a future paper.

  \ifanon\else
  \paragraph*{Acknowledgements}
  The second named author is especially grateful to Steve Awodey for his hospitality in August of 2023, during which the seeds of \emph{based Segal completeness} were planted in conversation with Reid Barton and Mathieu Anel. We thank Ulrik Buchholtz, Marcelo Fiore, Daniel Gratzer, Tom de Jong, Jonathan Weinberger, and Lingyuan Ye for their helpful advice.
  \fi
\end{xsect}
 
\nocite{phoa:1991}
\bibliographystyle{IEEEtran}
\bibliography{references/refs-bibtex}

\clearpage 
\appendix

\subsection{Omitted proofs}

\emph{Lemma~\ref{lem:fiore-lemma}:}
  Let $\mathcal{L}$ be a class of maps for which $\II$ is $\mathcal{L}$-local; if $A$ is $\OpenEmbeddings^*\mathcal{L}$-local, then $L\prn{A}$ is $\mathcal{L}$-local.
\begin{IEEEproof}
  Fix $f\colon X\to Y$ in $\mathcal{L}$. We have the following definitionally commuting square:
  \[ 
    \begin{tikzcd}[column sep=large]
      L\prn{A}\Sup{Y}
        \arrow[r,"\text{distributivity}"]
        \arrow[d,swap,"L\prn{A}\Sup{f}"]
      &
      \sum\Sub{\prn{\alpha:\II\Sup{Y}}}A\Sup{Y.\alpha}
      \arrow[d,"\prn{\II^f,A\Sup{f.-}}"]
      \\ 
      L\prn{A}\Sup{X}
        \arrow[r,swap,"\text{distributivity}"]
      &
      \sum\Sub{\prn{\alpha:\II\Sup{X}}}A\Sup{X.\alpha}
    \end{tikzcd}
  \]

  Above, for each $\alpha\colon Z\to \II$ we have written $Z.\alpha$ for the corresponding subobject of $Z$ under $\IsT{-}\colon \II\to\Prop$. To see that the western map is an isomorhpism, we will use the three-for-two property of isomorphisms. The northern and southern maps are the canonical isomorphisms that distribute products over sums. To see that the eastern map is an isomorphism, we note that its first component $\II^f\colon \II^Y\to \II^X$ is an isomorphism by our assumption that $\II$ is $\mathcal{L}$-local; that the second component is an isomorphism of $\II^f$ follows from our assumption that $A$ is $\OpenEmbeddings^*\mathcal{L}$-local, as we have the following pullback square
  \[
    \begin{tikzcd}
      X.\alpha 
        \arrow[r,hookrightarrow]
        \arrow[d,swap,"f.\alpha"]
        \arrow[dr,phantom,near start,"\lrcorner"]
      &
      X
        \arrow[d,"f"]
      \\
      Y.\alpha
        \arrow[r,hookrightarrow]
        \arrow[d]
        \arrow[dr,phantom,near start,"\lrcorner"]
      &
      Y \arrow[d,"\alpha"]
      \\ 
      \brc{1}
        \arrow[r,hookrightarrow]
      &
      \II
    \end{tikzcd}
  \]

  By definition, $f.\alpha$ lies in $\OpenEmbeddings^*\mathcal{L}$, so we are done.
\end{IEEEproof}

\emph{Theorem~\ref{thm:based-segal-complete-vs-well-complete}:}
Suppose that $\II$ is a consistent bounded distributive lattice satisfying the disjunction property and the Phoa principle, equipped with an internal sum structure that factors binary meets. Then a type $X$ is \emph{based} Segal complete if and only if it is \emph{based} Segal well-complete.
\begin{IEEEproof}
  We shall show that based Segal completeness implies based Segal well-completeness, as the other direction is trivial. We begin by applying Lemma~\ref{lem:fiore-lemma}, after which we need only show that $\II$ is based Segal complete (Lemma~\ref{thm:interval-is-based-segal-complete}), and that $X$ is local with respect to the restriction of the evident inclusion $\eta_j\colon \IsT{j}_\bot\hookrightarrow \II/j$ along any open embedding. For the latter, we fix an arbitrary function $\alpha\colon \II/j\to \II$ and consider the restricted embedding $\eta_j.\alpha\colon \IsT{j}_\bot.\alpha\hookrightarrow\II/j.\alpha$ induced by pullback:
  \[ 
    \begin{tikzcd}
      \IsT{j}_\bot.\alpha
        \arrow[d,hookrightarrow]
        \arrow[r,hookrightarrow,"\eta_j.\alpha"]
        \arrow[dr,phantom,very near start,"\lrcorner"]
      &
      \II/j.\alpha
        \arrow[r]
        \arrow[d,hookrightarrow]
        \arrow[dr,phantom,very near start,"\lrcorner"]
      &
      \brc{1}
        \arrow[d,hookrightarrow]
      \\ 
      \IsT{j}_\bot
        \arrow[r,hookrightarrow,swap,"\eta_j"]
      &
      \II/j
        \arrow[r,swap,"\alpha"] 
      &
      \II
    \end{tikzcd}
  \]

  By the Phoa principle, have $\alpha_{i} = \alpha_0\sqcup \beta_i$ where $\beta_i = i\sqcap \alpha_j$. Using the disjunction property, we therefore obtain an explicit description of $\eta_j.\alpha\colon \IsT{j}_\bot.\alpha\hookrightarrow\II/j.\alpha$:
  \begin{align*}
    \IsT{j}_\bot.\alpha &= \Compr{i\in\IsT{j}_\bot}{\IsT{\alpha_0}\lor\IsT{\beta_i}}
    \\ 
    \II/j.\alpha &= \Compr{i\in\II/j}{\IsT{\alpha_0}\lor\IsT{\beta_i}}
  \end{align*}

  We will render the above as pushout squares by first compting the conjunction $\IsT{\alpha_0}\land\IsT{\beta_i}$ for any $i\in\II/j$:
  \[ 
    \IsT{\alpha_0}\land\IsT{\beta_i}
    =
    \IsT{\alpha_0}\land\IsT{i}\land\IsT{\alpha_j}
    =
    \IsT{\alpha_0}\land\IsT{i}
  \]
  
  Furthermore, we can see that $\IsT{j}_\bot.\beta\subseteq\IsT{j}_\bot$ and $\II/j.\beta\subseteq\II/j$ are both precisely the subsingleton subsets containing only $1$ such that $\IsT{j}\times\IsT{\alpha_j}$ holds.
  Therefore, we have the following span of pushouts:
  \[ 
    \begin{tikzcd}
      \II/j\times\IsT{\alpha_0}
        \arrow[d,hookrightarrow]
      &
      \brc{1}\times\IsT{j}\times\IsT{\alpha_0} 
        \arrow[dr,phantom,very near end,"\ulcorner"]
        \arrow[dl,phantom,very near end,"\urcorner"]
        \arrow[r,hookrightarrow]
        \arrow[d,hookrightarrow]
        \arrow[l,hookrightarrow]
      &
      \IsT{j}_\bot\times\alpha_0  
        \arrow[d,hookrightarrow]
      \\
      \II/j.\alpha
      &
      \brc{1}\times\IsT{j}\times\IsT{\alpha_j}
        \arrow[r,hookrightarrow]
        \arrow[l,hookrightarrow]
      &
      \IsT{j}_\bot.\alpha
    \end{tikzcd}
  \] 

  By the universal property of the left-hand square, the following square is cartesian:
  \[ 
    \begin{tikzcd}
      X\Sup{\II/j.\alpha}
        \arrow[dr,phantom,very near start,"\lrcorner"]
        \arrow[r,hookrightarrow]
        \arrow[d,hookrightarrow]
      &
      X\Sup{\II/j\times\IsT{\alpha_0}}
        \arrow[d,hookrightarrow]
      \\ 
      X\Sup{\brc{1}\times\IsT{j}\times\IsT{\alpha_j}}
        \arrow[r,hookrightarrow]
      &
      X\Sup{\brc{1}\times\IsT{j}\times\IsT{\alpha_0}}
    \end{tikzcd}
  \] 

  We have assumed that $X$ is $\eta_j$ local, so the restriction map $X\Sup{\eta_j\times\IsT{\alpha_0}}\colon X\Sup{\II/j\times\IsT{\alpha_0}}\to X\Sup{\IsT{j}_\bot\times\IsT{\alpha_0}}$ is an isomorphism; thus we have the following pasting of cartesian squares:
  \[ 
    \begin{tikzcd}[cramped]
      X\Sup{\II/j.\alpha}
        \arrow[dr,phantom,near start,"\lrcorner"]
        \arrow[r]
        \arrow[d]
      &
      X\Sup{\II/j\times\IsT{\alpha_0}}
        \arrow[dr,phantom,near start,"\lrcorner"]
        \arrow[d]
        \arrow[r,"X\Sup{\eta_j\times\IsT{\alpha_0}}"]
      &
      X\Sup{\IsT{j}_\bot\times\IsT{\alpha_0}}
        \arrow[d]
      \\ 
      X\Sup{\brc{1}\times\IsT{j}\times\IsT{\alpha_j}}
        \arrow[r]
      &
      X\Sup{\brc{1}\times\IsT{j}\times\IsT{\alpha_0}}
        \arrow[r,equal]
      &
      X\Sup{\brc{1}\times\IsT{j}\times\IsT{\alpha_0}}
    \end{tikzcd}
  \] 

  By definition, $X\Sup{\IsT{j}_\bot.\alpha}$ is the pullback of the same outer co-span and the gap isomorphism $X\Sup{\II/j_\bot.\alpha}\to X\Sup{\IsT{j}_\bot.\alpha}$ is precisely the restriction map $X\Sup{\eta_j.\alpha}$.
\end{IEEEproof}
 
\end{document}